\documentclass[11pt,letterpaper,onecolumn]{article}
\usepackage[margin=1in]{geometry}
\usepackage{amsmath,amssymb}
\usepackage{authblk}
\usepackage[numbers,sort&compress]{natbib}
\usepackage{booktabs}
\usepackage{placeins}
\usepackage[title]{appendix}
\usepackage{graphicx}
\usepackage[english]{babel}
\usepackage{bm}
\usepackage{amsthm}
\usepackage{microtype}
\usepackage{xurl}
\usepackage[hidelinks]{hyperref}

\mathchardef\mhyphen="2D
\newcommand*\diff{\mathop{}\!\mathrm{d}}

\def\fp#1#2{\frac{\partial #1}{\partial #2}}
\def\fd#1#2{\frac{\delta #1}{\delta #2}}
\def\l#1{\left#1}
\def\r#1{\right#1}

\def\d{\mathrm{d}}
\def\Tr{\text{Tr}}
\newtheorem{theorem}{Theorem}

\newtheorem{proposition}[theorem]{Proposition}

\newcommand{\Rmnum}[1]{\uppercase\expandafter{\romannumeral #1\relax}}
\newcommand{\rmnum}[1]{\lowercase\expandafter{\romannumeral #1\relax}}
\allowdisplaybreaks[1]

\begin{document}
\title{Variational learning of energy-stable macroscopic flow equations from microscopic laws}
\author[1]{Pei Ge}
\author[2,3]{Huan Lei\thanks{Corresponding author: \href{mailto:leihuan@msu.edu}{leihuan@msu.edu}.}}
\affil[1]{Department of Mathematical Sciences, Worcester Polytechnic Institute, Worcester, MA 01609}
\affil[2]{Department of Computational Mathematics, Science \& Engineering, Michigan State University, East Lansing, MI 48824, USA}
\affil[3]{Department of Statistics \& Probability, Michigan State University, East Lansing, MI 48824, USA}
\date{}
\maketitle

\begin{abstract}
We present a generalized variational approach for constructing energy-stable macroscopic partial differential equations (PDEs) directly from microscopic physical laws. Rather than fitting the PDE evolution for prescribed variables from macroscopic trajectories, we construct an extendable variational structure by adjoining auxiliary variables to seek autonomous reduced dynamics in an enlarged state space. By defining these variables as micro--macro encoders with explicit microscopic counterparts, we cast the learning problem as the sample-wise joint identification of the reduced state representation and the variational ingredients from time-discrete data, enabling systematic law-level transfer across scales. The resulting PDEs preserve frame indifference, physical constraints, and energy stability by construction, and admit a fluctuation--dissipation-consistent stochastic extension. Numerical results for the non-Newtonian hydrodynamics of polymeric fluids show that the learned models capture the viscoelastic responses induced by nonlinear and heterogeneous molecular interactions beyond empirical closures.
\end{abstract}

\section{Introduction}
Multiscale modeling seeks tractable macroscopic descriptions of collective behavior emerging from microscopic dynamics \cite{More_Anderson_Science_1972}. When clear scale separation exists, asymptotic reduction can often guide the choice of resolved variables and derive their effective dynamics. Micro--macro frameworks such as the heterogeneous multiscale method \cite{HMM_2003} and the equation-free approach \cite{Equation_Free_2003} exploit such separation by using short microscopic simulations to infer macroscopic dynamics on the fly. However, for systems without clear scale separation, the situation is fundamentally less settled. While macroscopic reduced models are often formulated as autonomous partial differential equations (PDEs) for selected field variables, such as density, velocity, temperature, or other structural variables, the prescribed set of variables may not admit such an autonomous closure. Classical moment approaches \cite{Grad_CPAM_1949,Levermore_entropy_closure_JSP_1996} enlarge the resolved state with higher-order moments, while the adequacy of the prescribed moment set remains problem dependent.  For such systems, an adequate set of resolved variables and closure relations are generally unknown \emph{a priori}, and transferring the microscopic physical laws to a structurally admissible reduced description remains a central challenge.

Recent data-driven approaches, including sparse-regression \cite{Schaeffer_Osher_2013, SINDy_2016, PDE_FIND_2017, Weak_SINDy_PDE_2021} and neural network (NN)-based \cite{PDE_Net_2018} methods, have advanced the inference of reduced dynamics from spatiotemporal observations. Notably, machine learning (ML)-based moment closures have enabled kinetic-to-hydrodynamic reduction while preserving conservation laws \cite{Han_Ma_PNAS_2019}. 
Beyond closure inference, macroscopic models must also preserve the underlying structure, such as frame indifference, thermodynamic admissibility, and energy stability. Rooted in Onsager's reciprocal relations \cite{Onsager_Phys_Rev_1,Onsager_Phys_Rev_2}, variational and thermodynamic formulations \cite{morrison1986paradigm, Grmela_Ottinger_1997_1, Ottinger_beyond_2005, beris1994thermodynamics, Lin_Liu_Zhang_2005, giga_variational_2018} encode 
the coupling of reversible transport and irreversible dissipation, yielding macroscopic models with energy stability by construction. Recent ML approaches \cite{YuTianELi2021,HuangHeChemReina2022,zhang_gfinns_2022,Masi_Stefanou_CMAME_2022,chen_constructing_2023,park_tlasdi_2024, nieto_simavilla_hammering_2025, shiratori_gimlet_2025} identify the corresponding variational ingredients from macroscopic evolution data. 
Despite these advances, two fundamental difficulties remain for multiscale modeling. Existing learning formulations are generally based on pre-selected resolved variables \cite{huang_statistical-physics-informed_2025}. This assumes that the chosen variables admit autonomous dynamics, which may not be guaranteed.  
Moreover, the reduced dynamics is often inferred from trajectories generated by microscopic simulations. 
Such trajectory-based learning accesses microscopic laws only through their projected macroscopic responses, rather than directly exploiting their explicit analytical form.
Systematic law-level transfer in data-driven multiscale closure with physical constraints remains largely open.

In this work, we develop a generalized variational learning framework for constructing energy-stable macroscopic PDE models. 
The central novelty is the joint identification of the reduced state representation and its variational closure by directly exploiting the microscopic laws, rather than fitting the evolution in PDE form for prescribed variables from macroscopic trajectories. 
The framework generalizes an extendable variational structure \cite{Ottinger_beyond_2005, beris1994thermodynamics, giga_variational_2018}, adjoining a set of auxiliary field variables with explicit microscopic counterparts, to adaptively seek closed and structurally admissible reduced dynamics in an enlarged state space. Crucially, the auxiliary variables serve as physically interpretable micro--macro encoders, whose macroscopic evolution follows analytically from the microscopic dynamics. This analytical correspondence enables the microscopic laws to be directly transferred to the macroscopic closure. Consequently, the reduced state representation and variational ingredients are identified sample-wise from time-discrete data and parameterized by symmetry-preserving NNs. Upon variation, the learned structure yields autonomous PDEs satisfying physical and symmetry constraints, and enables stable numerical discretization and long-time simulation. 

We illustrate the framework through the non-Newtonian hydrodynamics of polymeric fluids, a canonical multiscale problem where macroscopic stress and transport depend on microscopic molecular interactions \cite{Owens_Phillips_2002,Bird_Curtiss_book_vol_2,REN_E_HMM_complex_fluid_2005}. The hydrodynamic equations contain an unclosed polymer stress coupled to constitutive equations for suitable structural variables. Classical models \cite{Larson88} typically prescribe low-dimensional moments of the polymer configuration distribution, such as end-to-end conformation tensors, together with empirical forms for stress, objective tensor derivatives, and relaxation dynamics. While these models capture qualitative viscoelastic responses, their molecular fidelity may deteriorate with nonlinear or heterogeneous molecular interactions. Recent data-driven approaches \cite{Lennon_PNAS_2023,Young_SIMPLE_2023} have learned constitutive models from rheological data and scattering-based microstructural trajectories. 
Within the present framework, we introduce auxiliary variables as generalized conformation tensors, interpreted as nonlinear features of the polymer configuration distribution with explicit microscopic counterparts. These tensors and variational ingredients are identified from time-discrete samples using the microscopic laws. The constructed variational structure generates the polymer stress and constitutive dynamics jointly, rather than fitting them separately at the PDE level as in \cite{Lei_Wu_E_2020,Fang_Ge_JML_2022}.  The resulting model captures molecular interaction effects beyond empirical closures, while preserving frame indifference, energy stability, and the positive definiteness of the conformation tensors, and further enables a stochastic extension with a consistent fluctuation--dissipation relation.
Although demonstrated for polymeric fluids, the framework extends in principle to other multiscale problems, including kinetic-to-hydrodynamic reduction \cite{Han_Ma_PNAS_2019}.

\section{Preliminary}
\subsection{Microscopic description}
We consider a polymer solution consisting of $N_m$ molecules, with each polymer represented by a bead--spring model containing $N$ particles. 
Let $\bm q\in\mathbb{R}^{N\times 3}$ denote the configuration of a molecule and let $\bm r=\bm K\bm q$  
collect its $N_b$ bond vectors, where $\bm r_n \in\mathbb{R}^{3\times 1}$ is the $n-$th bond vector and $\bm K\in\mathbb{R}^{N_b\times N}$ is the connectivity matrix determined by the molecular structure. The intramolecular potential of the polymer system is
\begin{equation}
V_m=\sum_{m=1}^{N_m}V\!\left(\bm r^{(m)}\right),
\qquad
V(\bm r)=\sum_{n=1}^{N_b}V_b\!\left(|\bm r_n|\right),
\label{eq:micro_potential}
\end{equation}
where $V_b$ denotes the bond potential and the superscript indexes the polymer molecules. 
In principle, the molecular architecture $\bm K$ and bond interactions $V_b$,
together with the coupling to the surrounding solvent
and thermal fluctuations, determine the microscopic dynamics of the polymer
solution. In practice, directly resolving all polymer and solvent degrees of freedom over
hydrodynamic scales could become computationally prohibitive. This motivates the macroscopic reduced model based on various empirical closures \cite{Doi_Theory_Polymer_Book, Owens_Phillips_2002}.

\subsection{Macroscopic description}
At the continuum level, the incompressible hydrodynamic equations take the form
\begin{equation}
\begin{split}
\nabla\cdot\bm u &=0 \\
\rho\left(\frac{\partial\bm u}{\partial t}
+\bm u\cdot\nabla\bm u\right)
&=
-\nabla p+\nabla\cdot(\bm\tau_s+\bm\tau_p)+\bm g,    
\end{split}
\label{eq:momentum_transport_close}
\end{equation}
where $\rho$, $\bm u$, $p$, and $\bm g$ denote the fluid density,
velocity, pressure, and external body force, respectively. $\bm \tau_{s}$  is the solvent stress tensor and takes the closed Newtonian form 
$\bm \tau_s  = \eta(\nabla \bm u + \nabla \bm u^\top)$ with shear viscosity $\eta$. However, a closure gap arises from the polymer stress $\bm \tau_{p}$. In principle, $\bm\tau_{p}$ can be determined from the microscopic interactions in Eq. \eqref{eq:micro_potential}. 
Let $f(\bm r,\bm x,t)$ denote the local
probability density function (PDF) of the polymer configuration, and let
$\langle\cdot\rangle_{\bm x,t}$ denote the corresponding ensemble average with respect to $f$. For notational simplicity, we suppress the subscript \((\bm x,t)\) hereafter. $\bm\tau_p$ takes the microscopic form
\begin{equation}
\bm\tau_p(\bm x,t)
=
n_p
\left\langle
\sum_{n=1}^{N_b}
\bm r_n\otimes\nabla_{\bm r_n}V(\bm r) 
\right\rangle - n_p N_b k_BT \bm I,
\label{eq:micro_polymer_stress}
\end{equation}
where $n_p$ is the polymer number density and is assumed spatially homogeneous as a constant.  

To close the macroscopic model in Eq.~\eqref{eq:momentum_transport_close},  a common approach  is to approximate $\bm \tau_{\rm p}$ in terms of a pre-selected conformation tensor $\bm c$ along with its constitutive dynamics, i.e., 
\begin{equation}
\bm \tau_{p}  = \bm G(\bm c) \qquad \frac{\mathcal{D}{\bm c}}{\mathcal{D} t} = \bm H(\bm c).
\label{eq:empirical_stress_dynamics}
\end{equation}
For simple dumbbell-shaped molecules with $N_b = 1$, a standard choice is the conformation tensor $\bm c = \left\langle \bm r\bm r^\top\right\rangle$. $\bm G$ and $\bm H$ represent the stress and reduced dynamics. Their constitutive forms are often determined by different (e.g., linear or mean-field) approximations while preserving the symmetry constraints, such as the Hookean \cite{Oldroyd_Wilson_PRSLA_1950}, Giesekus \cite{Giesekus_JNNFM_1982}, FENE-P \cite{Peterlin_Polymer_Science_1966}, and FENE-S \cite{Yu_Du_mms_2005}. Notably, $\frac{\mathcal{D}{\bm c}}{\mathcal{D} t}$ denotes an objective tensor derivative, which also needs to satisfy the frame-indifference constraint. This constraint does not uniquely determine its form; common choices include upper-convected \cite{Oldroyd_Wilson_PRSLA_1950} and co-rotational \cite{Zaremba_1903}.

Despite the broad applications, Eq. \eqref{eq:empirical_stress_dynamics} is based on a prescribed reduced state and specific constitutive assumptions. For general molecular structures and interactions, neither an adequate reduced state nor its constitutive structure is known \emph{a priori}.
The problem addressed here is to jointly identify the reduced state and an extendable variational structure directly from the microscopic laws, by
which the stress, objective transport, and reduced dynamics are generated rather than fitting the individual PDE forms for pre-selected variables.

\section{Learnable extendable variational non-Newtonian hydrodynamic model}

\subsection{Extendable admissible variational structure}
We consider the open isothermal incompressible hydrodynamic model in Eq.~\eqref{eq:momentum_transport_close} and embed it in an enlarged state space. Let $\bm\chi = (\bm M, \bm c_1, \cdots, \bm c_{N_c})$ represent a set of macroscopic field variables where $\bm M = \rho \bm u$ is the fluid momentum with $\nabla \cdot \bm M = 0$ and
all functional derivatives with respect to $\bm M$ are constrained variations on the divergence-free subspace. $\bm c_i\in \mathbb{S}_{+}^3$ is a set of 
generalized second-order conformation tensors representing learned nonlinear features of the polymer configuration PDF $f(\bm r, \bm x, t)$.
Their microscopic definitions will be identified jointly with the variational closure below.  

To construct the reduced dynamics, we represent the evolution of a general observable $F[\bm\chi]$ as the coupling of a reversible and an irreversible process taking a generalized energy variational form \cite{Ottinger_beyond_2005, beris1994thermodynamics, giga_variational_2018}, i.e., 
\begin{equation}
\dot{F}= \{F, \mathcal{H}\} + (F,\mathcal{H})_{-},
\label{eq:general_bracket_variation}
\end{equation}
where $\{\cdot, \cdot\}$ is the skew-symmetric Poisson bracket that generates the reversible transport  and $(\cdot, \cdot)_{-}$ is a non-positive dissipative bracket that governs the irreversible energy dissipation and molecular relaxation, i.e., 
\begin{equation}
\{F, G\} = -\{G, F\} \qquad (F, F)_{-} \le 0.
\label{eq:bracket_property}
\end{equation}
$\mathcal{H}$ is the free energy functional consisting of the kinetic energy and the polymer entropy, i.e., 
\begin{equation}
\mathcal{H}[\bm\chi] = \int_{\Omega} \left [\frac{\vert \bm M \vert^2}{2\rho} - s(\bm c_1, \cdots, \bm c_{N_c}) \right]\diff \bm x,
\label{eq:free_energy}
\end{equation}
where $s$ is the macroscopic entropy density and $-s$ the configuration free energy density with the detailed form to be learned. By imposing 
Eq. \eqref{eq:bracket_property}, the macroscopic model in Eq.~\eqref{eq:general_bracket_variation} ensures the free energy dissipation $\frac{\diff \mathcal{H}}{\diff t} \le 0$ in the absence of external body force. An external body force $\bm g$ is treated as prescribed
in the momentum balance without modifying the variational structures.

For two general observables $F[\bm \chi]$ and $G[\bm \chi]$, the Poisson bracket takes the extendable form
\begin{equation}
\{F,G\} = \int_{\Omega}\! \bm M\cdot \left[ (G_{M}\cdot\nabla)F_{M} - (F_{M}\cdot\nabla)G_{M} \right] \diff \bm x + \sum_{i = 1}^{N_c}\! \int_{\Omega}\! \left[F_i: \mathcal{T}_{G_{M}}\bm c_i - G_i: {\mathcal{T}}_{F_{M}}\bm c_i \right] \diff \bm x,
\label{eq:extendable_poisson_bracket}
\end{equation}
where $F_M := \delta F/\delta \bm M$  under constraint $\nabla \cdot \bm M = 0$, $F_{i} := \delta F/\delta \bm c_{i}$ and $\mathcal{T}_{\bm v} \bm c:= -\bm v\cdot\nabla\bm c +(\nabla\bm v)\bm c+\bm c(\nabla\bm v)^\top$.
By Ref. \cite{marsden_coadjoint_1983}, Eq. \eqref{eq:extendable_poisson_bracket} satisfies the Jacobi identity and forms the incompressible hydrodynamic transport. 

To model the constitutive dynamics, we construct the non-positive dissipative bracket consisting of both symmetric and skew-symmetric parts. The symmetric part governs the free energy dissipation and takes the extendable form 
\begin{equation}
(F, G)_{-}^s = - \int_{\Omega} \frac{\eta}{2} \l({\bm D} F_{M}\r): \l({\bm D} G_{M}\r) \diff \bm x -\sum_{i,j=1}^{N_c} \int_{\Omega} F_{i}:\mathcal{M}_{ij}:G_{j} \diff \bm x,
\label{eq:dissipative_symmetry}
\end{equation}
where $\bm D \bm v := \nabla \bm v + \nabla {\bm v}^{\top}$. The first part represents the dissipation due to the Newtonian stress with viscosity $\eta$. The second part represents the polymer molecule relaxation, where $\mathcal{M}_{ij}(\bm c_1, \cdots, \bm c_{N_c}) \in \mathbb{R}^{3\times3\times3\times3}$ represents the conformation-dependent friction matrices coupling among the individual features. The block operator 
$\bm {\mathcal{M}} = [\mathcal{M}]_{i,j=1}^{N_c}$ is self-adjoint, frame-indifferent, with $\bm {\mathcal{M}} \succeq 0$. Its detailed form will be learned. 

In addition, we introduce a skew-symmetric part 
\begin{equation}
(F, G)_{-}^a = \sum_{i=1}^{N_c} \int_{\Omega}
\left[ F_{i} : (\nabla G_{M}:\mathcal{E}_{i}) 
 -  G_{i} : (\nabla F_{M}:\mathcal{E}_{i}) \right] \diff \bm x, 
\label{eq:dissipative_skew}
\end{equation}
where $\mathcal{E}_{i}(\bm c_1, \cdots, \bm c_{N_c}) \in \mathbb{R}^{3\times3\times3\times3}$ is frame-indifferent and will be learned from microscopic models. 
By skew symmetry, $(F,F)_{-}^{a}=0$, so this term does not contribute to the free energy dissipation. With $\mathcal{E}_{i} = 0$, the transport recovers the upper-convected tensor derivative used in common constitutive models. As shown later, this skew-symmetric term with non-zero $\mathcal{E}_{i}$  enlarges the class of admissible frame-indifferent tensor dynamics by specifying the additional nonlinear coupling between the flow field $\bm u$ and the individual molecular features $\bm c_{i}$, and is crucial for constructing their generalized objective tensor derivative $\mathcal{D} \bm c_{i}/\mathcal{D} t$ directly from their microscopic counterparts with clear physical interpretation.

By specifying the proper form of the macro-features $\{\bm c_{i}\}_{i=1}^{N_c}$, entropy density $s$, friction operator $\bm {\mathcal{M}}$, and skew-symmetric coupling $\mathcal{E}$, Eq. \eqref{eq:general_bracket_variation} provides the foundation to generate the admissible hydrodynamic PDE model, including the closure for the polymer stress and conformation dynamics in Eq. \eqref{eq:empirical_stress_dynamics}.  
Before we proceed, we note that the present variational structure is constructed for the isothermal open fluid system in Eq. \eqref{eq:momentum_transport_close}. For general closed systems, 
we can further split the single free energy formulation into separate energy $E$ and entropy $S$, i.e., $\dot{F}= \{F, E\} - (F, S)_{-}$ with the degeneracy condition $\{F, S\}\equiv 0$ and $(F, E)_{-} \equiv 0$ for a general $F$; see the GENERIC form \cite{Ottinger_Grmela_1997_2}. This ensures energy conservation
$\diff E/\diff t \equiv 0$ and entropy production $\diff S/\diff t \ge 0$. Appendix~\ref{app:closed_system} gives the closed-system formulation. 

\subsection{Extendable hydrodynamic PDE form} By taking $F = \bm\chi(\bm x, t)$ componentwise, the present variational form in Eq.~\eqref{eq:general_bracket_variation}, along with the free energy functional and brackets in Eqs. \eqref{eq:free_energy}--\eqref{eq:dissipative_skew}, generates the intrinsic hydrodynamic and constitutive dynamics. By including the prescribed external force $\bm g$, we have the following extendable PDE class 
\begin{equation}
\begin{split}
\nabla\cdot\bm u &=0 \\
\rho\left(\partial_t\bm u
+\bm u\cdot\nabla\bm u\right)
&=
-\nabla p+ \eta \Delta \bm u + \nabla\cdot \bm \tau_p + \bm g\\
\partial_t \bm {c}_{i}
+ \bm u \cdot \nabla \bm {c}_{i} - \bm \kappa : \tilde{\mathcal{E}}_{i} & =  \sum_{j=1}^{N_c} \mathcal{M}_{ij} :\frac{\partial s}{\partial \bm{c}_{j}} \\
\bm\tau_p &= - \sum_{i=1}^{N_c} \tilde{\mathcal{E}}_{i}: \frac{\partial s}{\partial \bm c_{i}}, 
\end{split}
\label{eq:variational_PDE_form}
\end{equation}
where $\bm \kappa = \nabla \bm u^{\top}$, $(\widetilde{\mathcal E}_i)_{klmn}
= (\mathcal E_i)_{klmn} + (c_i)_{km}\delta_{ln}
+(c_i)_{kn}\delta_{lm}$,
and $-\nabla p$ arises from imposing $\nabla \cdot \bm u = 0$ through the Leray--Hodge projection onto the divergence-free space. Formally, $\mathbb{P} \bm v = \bm v - \nabla \Delta^{-1} (\nabla \cdot \bm v)$ and the variation with respect to $\bm M$ is understood as $\mathbb{P} (\delta F/\delta \bm M)$. Appendix~\ref{app:open_system} derives the momentum and conformation equations from the brackets.

Unlike the conventional constitutive modeling form in Eq. \eqref{eq:empirical_stress_dynamics}, the polymer stress and conformation dynamics in Eq. \eqref{eq:variational_PDE_form} are generated jointly by the variational structure rather than specified separately at the PDE level. 
To close this PDE class, we will need to construct symmetry-preserving representations of $s$, $\mathcal M$, and $\mathcal E$, together with microscopic representations of the generalized conformation variables $\{\bm c_i\}_{i=1}^{N_c}$. 
These representations specify an admissible model class whose specific forms will be subsequently identified from the microscopic models. 

\subsection{Structure-preserving neural-network representation} We construct the macroscopic conformation tensors as the ensemble average of the microscale polymer configuration features, i.e., $\bm c_i = \left\langle \bm b_i(\bm r)\right\rangle$, where $\bm b_i$ is a learnable encoder of the polymer configuration. As a vector-valued encoder would yield the trivial feature $\left\langle \bm b_i\right\rangle \equiv 0$ due to the rotational symmetry, we choose $\bm b_i$ as second-order tensor-valued features taking the form
\begin{equation}
   \bm c_i = \left\langle \bm b_i(\bm r)\right\rangle \quad \bm b_i = \bm h_i \bm h_i^\top \quad \bm h_i = g_i(\bm q^{\ast}) \sum_{k=1}^{N-1} w_{ik} \bm r_k, \label{eq:encoder_b}
\end{equation} 
where $\bm q^{\ast} \in \mathbb{R}^{3N-6}$ is a translational- and rotational-invariant representation of the molecular configuration and $\{\bm r_k\}_{k=1}^{N-1}$ is a set of $N-1$ independent bond vectors selected from $\bm r$. $g:\mathbb{R}^{3N-6}\to\mathbb{R}^{N_c}$ is a NN to be learned. Although $\bm b_i$ and $\bm c_i$ are second-order tensors, they are not restricted to quadratic moments as in standard moment closures \cite{Levermore_entropy_closure_JSP_1996}, and encode  nonlinear configuration features through $g_i$. The encoders can further be normalized so that $\bm c_i$ takes a prescribed isotropic value at equilibrium, e.g., $\bm c_i^{\rm eq}=\bm I$. It is straightforward to verify the frame-indifference condition $\bm b_i(\bm r\mathcal{U}^\top) = \mathcal{U} \bm b_i \mathcal{U}^\top$, $\mathcal{U}\in SO(3)$.

The polymer entropy density $s$ is required to be frame-indifferent, concave, and maximized at the prescribed equilibrium state. For brevity,
we denote $\bm c=(\bm c_1,\ldots,\bm c_{N_c})$ and represent
\begin{equation}
\begin{split}
s(\bm c) &= \psi_s(\bm c) - \left .\frac{\partial \psi_s}{\partial \bm c}\right \vert_{\bm c^{\rm eq}} : \bm c \\
\psi_s(\bm c) &= -\psi_e(\bm c) + s_c(\bm c) \quad s_c(\bm c) = \sum_{i=1}^{N_c} \lambda_i \log {\det} \bm c_i,
\end{split}
\label{eq:entropy_density}
\end{equation}
where $\lambda_i > 0$. Here, $\psi_e$ represents the learned effective elastic free energy
contribution, while $s_c$ is the Casimir contribution. In particular,
$\mathcal S_c=\int_\Omega s_c\, \diff \bm x$ is invariant under the Poisson bracket in Eq.~\eqref{eq:extendable_poisson_bracket}, i.e.,
$\{F,\mathcal S_c\}\equiv0$. We represent
$\psi_e(\bm c)=\Phi_\theta(\bm x_c)$, where

\begin{equation*}
\bm x_c=
\left\{
{\rm Tr} (\bm c_i),\,
{\rm Tr}(\bm c_i^2),\,
{\rm Tr}(\bm c_i^3),\,
{\rm Tr}\!\left[(\bm c_i-\bm c_j)^2\right]
\right\}_{i,\,i<j}
\end{equation*}
 and $\Phi_\theta$ is a modified input-convex neural
network \cite{ICNN_Amos_Xu_2017} that is componentwise nondecreasing. Each component of $\bm x_c$ is rotationally
invariant and convex with respect to $\bm c$. 
Hence,  $\psi_e$ is convex and
frame-indifferent, so $\psi_s$ is concave and frame-indifferent, while the affine term ensures that $\bm c^{\rm eq}$ is a maximizer of $s$. 

The fourth-order tensors $\mathcal E_i(\bm c)$ and the blocks
$\mathcal M_{ij}(\bm c)$ of the friction operator must preserve frame
indifference and tensor symmetry through their action on second-order
tensors. For $\mathcal K=\mathcal E_i$ or $\mathcal M_{ij}$, we require
\begin{equation}
\begin{split}
\widetilde{\bm R:\mathcal K(\bm c)}
&=
\widetilde{\bm R}:\mathcal K(\widetilde{\bm c})
\qquad
\widetilde{\mathcal K(\bm c):\bm R}
=
\mathcal K(\widetilde{\bm c}):\widetilde{\bm R}
\\
[\bm R:\mathcal K(\bm c)]^\top &= \bm R:\mathcal K(\bm c)
\quad
[\mathcal K(\bm c):\bm R]^\top = \mathcal K(\bm c):\bm R,
\end{split}
\label{eq:fourth_order_tensor_symmetry}
\end{equation}
where $\bm R\in\mathbb R^{3\times 3}$, $\widetilde{\bm R}=\mathcal U\bm R\mathcal U^\top$,
$\widetilde{\bm c}=\mathcal U\bm c\mathcal U^\top$ and
$\mathcal U\in SO(3)$.

To impose these constraints, we introduce a set of symmetric, rotationally equivariant polynomial tensor basis functions
$\{\bm p_s(\bm c)\}_{s=1}^{N_p}$. Pairwise products of these tensors are
combined into the fourth-order blocks designed to enforce
Eq.~\eqref{eq:fourth_order_tensor_symmetry} by representing
$\mathcal E$ and $\mathcal M$ as 
\begin{equation}
\begin{split}
&\{\bm {P}_r\}_{r=1}^{N_r}
=
\left\{
\bm p_s\boxtimes\bm p_t
+
\bm p_t\boxtimes\bm p_s,
\bm p_s\otimes\bm p_t
\right\}_{s,t=1}^{N_p}
\\
&\mathcal E_i 
=
\sum_r \xi_{ri}\bm {P}_r, 
\Gamma_{ij}
=
\sum_r \zeta_{r,ij} \bm {P}_r,  
\boldsymbol{\mathcal M} = \boldsymbol\Gamma\boldsymbol\Gamma^\ast + \epsilon \bm{\mathcal{I}},
\end{split}
\label{eq:fourth_order_basis_E_M}
\end{equation}
where $(\bm p\boxtimes \bm q)_{klmn}= p_{km}q_{ln} + q_{kn}p_{lm} $ and $(\bm p\otimes \bm q)_{klmn}= p_{kl}q_{mn}$.  $\bm \xi_{r}(\bm z_c) \in \mathbb{R}^{N_c}$ and $\bm \zeta_r (\bm z_c) \in \mathbb{R}^{N_c \times N_c} $ are NNs of rotational
invariants $\bm z_c$, including the convex components $\bm x_c$ introduced above. $\bm\Gamma^\ast$ denotes the
adjoint of $\bm\Gamma$. The term $\epsilon \bm{\mathcal{I}}$ with $\epsilon > 0$ and $\mathcal{I}_{ij} = \delta_{ij}(\delta_{km}\delta_{ln} + \delta_{kn}\delta_{lm})$ is a regularization term for $\bm{\mathcal{M}} \succ 0$ (see Prop. \ref{prop:positive_definiteness}). The construction in
Eq.~\eqref{eq:fourth_order_basis_E_M} preserves frame indifference and
tensor symmetry of the conformation dynamics and polymer stress, while
ensuring non-negative free energy dissipation of the PDE class in
Eq.~\eqref{eq:variational_PDE_form}. Appendix~\ref{app:nn_representation} specifies the encoder, entropy, and tensor representations and their network architectures.

\subsection{Microscopic law informed learning} With the admissible NN representations by Eqs. \eqref{eq:encoder_b}, \eqref{eq:entropy_density} and \eqref{eq:fourth_order_basis_E_M}, we establish joint learning of encoders $\left\{\bm b_i(\bm r)\right\}_{i=1}^{N_c}$, entropy density $s(\bm c)$, friction operator $\bm{\mathcal{M}}(\bm c)$ and the coupling tensor $\mathcal{E}(\bm c)$ directly from the microscopic laws. 
Crucially, the micro--macro encoders $\bm b_i(\bm r)$  bridge the physical laws across scales. Although their forms are unknown \emph{a priori}, they define learnable microscopic observables whose evolution can be derived analytically from the microscopic polymer dynamics in terms of the configuration PDF $f(\bm r, \bm x, t)$, and therefore provide direct microscopic counterparts for the macroscopic dynamics of $\bm c_i = \langle \bm b_i(\bm r) \rangle$.

In this study, we consider the micro-model as the common overdamped Langevin dynamics with constant friction coefficient $\gamma$, while the present learning framework applies to more complex micro-models such as Zimm \cite{Zimm_model_JCP_1956} with heterogeneous hydrodynamic interactions. Accordingly, the evolution of $f(\bm r, \bm x, t)$ is governed by the Fokker-Planck equation and the dynamics of $\bm c_i$ takes the form 
\begin{equation}
\begin{aligned}
&\partial_t \bm {c}_{i}+ \bm u \cdot \nabla \bm {c}_{i} - \boldsymbol{\kappa}: \tilde{\mathcal{E}}_i^{\rm micro} = \bm H_i^{\rm micro}\qquad \tilde{\mathcal{E}}_i^{\rm micro} = \left\langle \sum_{j=1}^{N-1} \bm r_j \otimes \nabla_{\bm r_j}\otimes \bm b_i \right\rangle
\\
&\bm H_i^{\rm micro} = \frac{k_BT}{\gamma} \left\langle \sum_{j,k=1}^{N-1} A_{jk} \nabla_{\bm r_j} \cdot \nabla_{\bm r_k} \bm b_i \right\rangle  - \frac{1}{\gamma} \left\langle \sum_{j=1}^{N-1} \sum_{k=1}^{N_b} A_{jk} \nabla_{\bm r_k} V(\bm r) \cdot \nabla_{\bm r_j} \bm b_i \right\rangle,
\end{aligned}
\label{eq:micro_consitutive_model}
\end{equation}
where $\bm A = \bm K \bm K^\top$ depends on the bond connectivity. We can show that the left-hand-side of the first equation is frame-indifferent and therefore forms a valid objective tensor derivative $\mathcal{D} \bm c_i/\mathcal{D}t$ with clear physical interpretation. Essentially, $\tilde{\mathcal{E}}_i^{\rm micro}$ and $\bm H_i^{\rm micro}$ in Eq. \eqref{eq:micro_consitutive_model} provide the microscopic counterparts for the unclosed constitutive terms in Eq. \eqref{eq:variational_PDE_form}. By matching the micro-macro counterparts, we establish the joint learning of $\{\bm b_i\}_{i=1}^{N_c}$, $s$, $\mathcal{M}$ and $\mathcal{E}$ by minimizing the empirical loss
\begin{equation}
\mathcal{L} = \sum_{l=1}^{N_s} \left( \Vert \bm H(\bm c^{(l)}) - \bm H^{{\rm micro},(l)}\Vert^2 + \Vert \tilde{\mathcal{E}}(\bm c^{(l)}) - \tilde{\mathcal{E}}^{{\rm micro},(l)}\Vert^2   + \Vert \tau_p(\bm c^{(l)}) - \tau_p^{{\rm micro},(l)}\Vert^2\right) ,
\label{eq:training_loss}
\end{equation}
where $\bm H_i = \sum_{j=1}^{N_c} \mathcal{M}_{ij} :\frac{\partial s}{\partial \bm{c}_{j}}$, $\tau_p = - \sum_{i=1}^{N_c} \tilde{\mathcal{E}}_{i}: \frac{\partial s}{\partial \bm c_{i}}$ and $\tau_p^{\rm micro}$ is defined by Eq. \eqref{eq:micro_polymer_stress}. 
Although not explicit in Eq.~\eqref{eq:training_loss},  encoders $\bm b_i$ determine both $\bm c_i$ and the microscopic targets, and are learned jointly with $s$, $\mathcal M$, and $\mathcal E$ through the micro--macro matching in Eq. \eqref{eq:training_loss}. The index $l$ labels time-discrete configuration ensembles sampled from the microscopic PDF $f(\bm r,\bm x,t)$. 
All microscopic target terms $\langle\cdot\rangle$ are evaluated by Monte Carlo averages over discrete microscopic samples using the analytical microscopic laws, without requiring the time derivatives from macroscopic trajectories. Appendix~\ref{app:training} gives the microscopic correspondence, the frame-indifference proof, and the training and comparison-model specifications. 

\subsection{Structure-preserving properties}
With the embedded variational structure and the symmetry-preserving NN representations, the constructed PDE model satisfies the prescribed physical symmetries and energy-dissipation structure, while the conformation tensors $\bm c_i$ remain admissible under the coercivity condition below.
\begin{proposition}
\label{prop:positive_definiteness}
Assume that $\psi_e(\bm c)$, $\mathcal E_i(\bm c)$, and
$\mathcal M_{ij}(\bm c)$ are $C^1$ on
$(\mathbb S_{++}^3)^{N_c}$ and admit continuous extensions to
$(\mathbb S_{+}^3)^{N_c}$. Suppose that 
$\boldsymbol{\mathcal M}$ is uniformly coercive, i.e.,
$
\bm A: \bm {\mathcal M}:\bm A
\ge
\epsilon \|\bm A\|_F^2,    
$
for $\bm A\in (\mathbb S^3)^{N_c}$ and some $\epsilon>0$. If $\bm c_i(0)\succ0$ and, along a
material trajectory, the conformation tensors and velocity gradient remain
bounded on a finite interval $[0,T)$, then there exists $\delta_T>0$ such that
\begin{equation}
\bm c_i(t)\succeq\delta_T\bm I,
\qquad
0\le t<T,\quad i=1,\ldots,N_c.
\end{equation}
\end{proposition}
Furthermore, the variational form enables the development of structure-preserving discretizations by adapting the energy-stable \cite{zhao2017numerical,shen2018scalar,Shen_Xu_Yang_SIAM_Review_2019} and positivity-preserving \cite{BALCI_Thomeses_Symm_JNNFM_2011,FATTAL200523,VAITHIANATHAN20063} numerical schemes. Also, it enables us to construct mesoscopic stochastic models by introducing coherent noise terms arising from thermal fluctuations.
For a mesoscopic state representing a finite number of polymers under constant flow velocity gradient $\bm \kappa$, 
the It\^o stochastic dynamics of $\bm c = (\bm c_1, \cdots, \bm c_{N_c})$ takes the form
\begin{equation}
\diff \bm c
=
\left[
\bm \kappa:\tilde{\mathcal{E}}
+
\boldsymbol{\mathcal M}\nabla_{\bm c}s
+
\beta^{-1}\,\nabla_{\bm c}\!\cdot
\boldsymbol{\mathcal M}
\right]\diff t
+
\bm{\mathcal B}\, \diff \bm W_t,
\label{eq:stochastic_model}
\end{equation}
where $\bm{\mathcal B}(\bm c)\bm{\mathcal B}(\bm c)^{\top}=2\beta^{-1} \boldsymbol{\mathcal M}(\bm c)$ and $\beta^{-1} = k_BT/V_c$ is the thermal energy normalized by a characteristic volume representing the mesoscopic state.  As shown later, the stochastic model in Eq.~\eqref{eq:stochastic_model} captures the polymer coil--stretch crossover under elongation flow and the associated fluctuation peak near the transition regime similar to experimental studies \cite{Perkins_Chu_Single_Science_1997}. The proof of Proposition~\ref{prop:positive_definiteness} is given in Appendix~\ref{app:properties}; Appendix~\ref{app:discretization} describes the structure-preserving discretizations.  

\section{Numerical results}

\subsection{Molecular fidelity beyond empirical closure} We examine the viscoelastic responses predicted by the macroscopic model in Eq. \eqref{eq:variational_PDE_form} constructed in comparison with the full microscopic model. On the microscale, the bond potential $V_b$ in Eq. \eqref{eq:micro_potential} takes the finite extensible nonlinear elastic (FENE) \cite{Warner_IECF_1972} form. Specifically, we consider both a chain-shaped and a star-shaped molecular structure with the same bead number $N = 7$ (with $N_b = 6$) and the same potential $V_b$. The full system consists of $N_m = 108000$ molecules in a $60\times100\times60$ domain (reduced unit) with periodic boundary conditions; the molecular interactions and simulation parameters are specified in Appendix~\ref{app:micro_model}. 

\begin{figure}[tbp]
\centering
\includegraphics[width=0.78\textwidth]{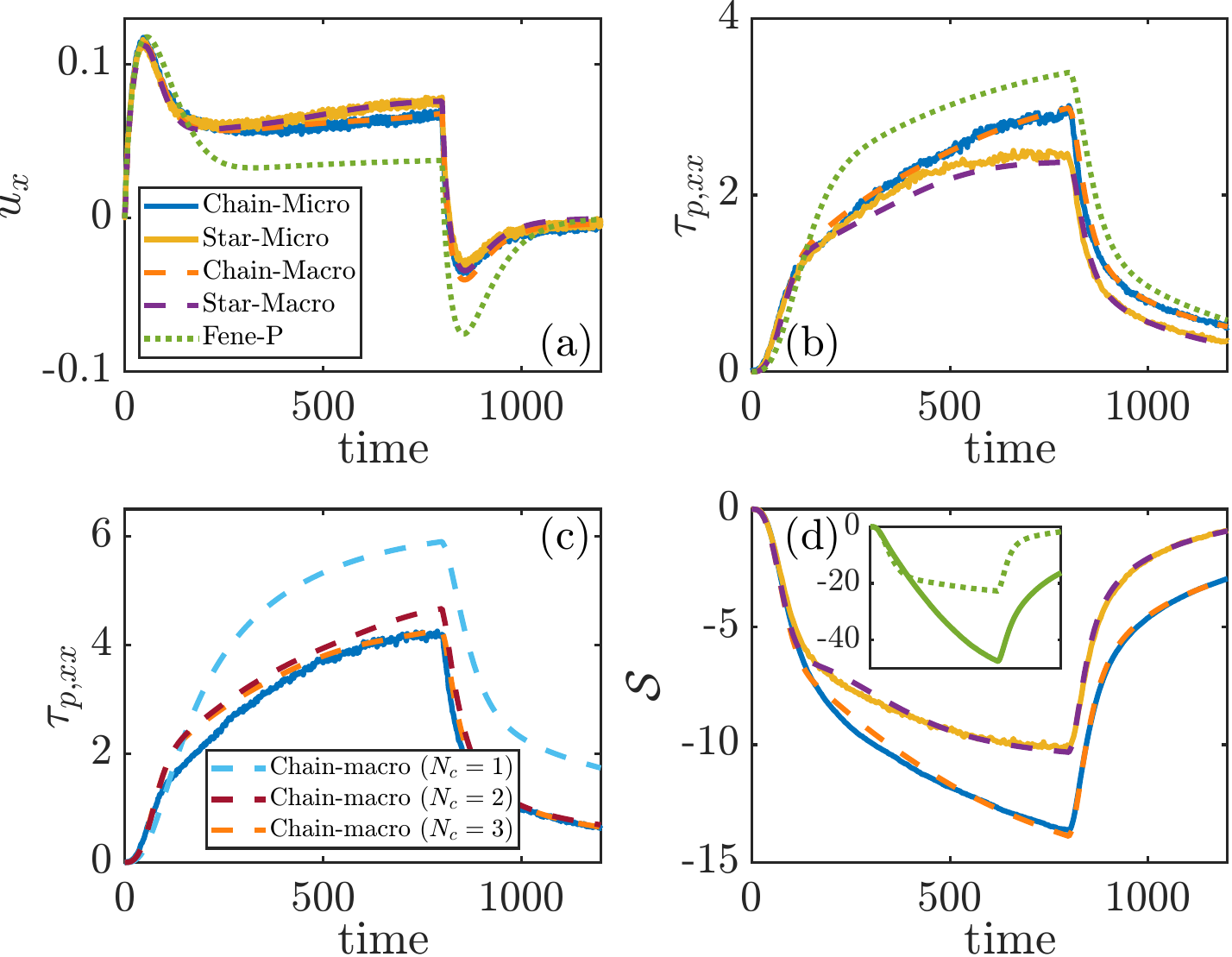}
\caption{The reverse Poiseuille flow of chain$\mhyphen$ and star$\mhyphen$shaped polymer suspensions with the body force $\bm g$ removed at $t=800$. (a) The velocity $u_x(y,t)$ at $y = 6$. (b) The polymer stress $\tau_{p,xx}(y,t)$ at $y=6$ (c) The polymer stress $\tau_{p,xx}(y,t)$ at $y = 1$. (d) Entropy $\mathcal{S} = \int_{\Omega} s(\bm c_1, \bm c_2,\bm c_3) \diff \bm x $. The inset plot shows the FENE-P form of $\int_{\Omega} s_\text{FENE-P}(\bm c_1) \diff \bm x $ evaluated on the conformation tensors from the microscopic chain model (solid) and the macroscopic FENE-P model (dashed).}
\label{fig:rpf_accuracy}
\end{figure}

On the macroscale, we simulate the reverse Poiseuille flow by applying opposite body force $\bm g=(\pm g_x, 0,0)$ on each half of the domain along the $y\mhyphen$direction with $g_x = 0.016$ starting from $t = 0$, and we set $g_x = 0$ for $t > 800$. Fig. \ref{fig:rpf_accuracy} shows the evolution of the velocity $u_x(y, t)$ and the stress $\tau_{p,xx}(y,t)$ for both the chain$\mhyphen$ and star$\mhyphen$shaped polymer solutions. Although the two systems are governed by the same potential $V_b$, their molecular structure leads to different viscoelastic responses. The conventional FENE-P closure \cite{Peterlin_Polymer_Science_1966,Bird_Doston_JNNFM_1980} using a
prescribed single conformation tensor is insufficient to capture such structure-dependent responses. In contrast, the predictions of the present model using  $3$ jointly learned variables $(\bm c_1, \bm c_2, \bm c_3)$ show good agreement with the full model results for both molecular structures. 

Fig. \ref{fig:rpf_accuracy}(c) shows the prediction of the stress development from the present model using different numbers of auxiliary variables. In particular, by using only one variable $\bm c_1$ (i.e., $N_c = 1$), the prediction also shows apparent deviations from the micro-model result. This indicates that the accurate learning of an empirical closure using one conformation tensor may not be sufficient for this system. Conversely, the present approach enables us to adaptively seek the closure in an enlarged space through the joint learning framework. 
Furthermore, the explicit variational form provides an energetic interpretation of the distinct flow responses. As shown in Fig. \ref{fig:rpf_accuracy}(d), the chain polymer undergoes a larger excursion in conformation space and reaches a lower entropy than the star polymer, which is consistent with larger elastic energy storage and a stronger viscoelastic response.
For both systems, the macroscopic entropy agrees with the same potential function on conformation tensors obtained from the microscopic configurations, whereas the FENE-P model exhibits apparent deviations. By learning an explicit entropy density \(s(\bm c_1,\bm c_2,\bm c_3)\) directly from microscopic physical laws, the present model naturally encodes such structural mechanics without human intervention.

\subsection{Model stability and robustness} Next, we examine the stability and robustness of the present model beyond the densely sampled training regime. Specifically, we establish long-time simulations of the reverse Poiseuille flow by imposing the external body force $\bm g$ up to $t = 2000$. The prolonged forcing drives the polymer configurations to larger deformation regimes that are not fully represented by the training data.

To isolate the roles of the variational structure and microscopic-law-informed
learning, we construct three additional macro-models using the same field variables
$(\bm c_1,\bm c_2,\bm c_3)$: (\rmnum{1}) The ``PDE'' model directly learns stress and constitutive dynamics in Eq. \eqref{eq:empirical_stress_dynamics} from time-series trajectories; (\rmnum{2}) ``VAR'' retains the variational form in Eq. \eqref{eq:variational_PDE_form} with
$\mathcal E_i=0$, and learns $s$ and $\mathcal{M}$ using time-series samples; (\rmnum{3})``PDE-Law'' learns the unclosed terms in Eq. \eqref{eq:empirical_stress_dynamics} by matching their analytical microscopic counterparts in Eq. \eqref{eq:micro_consitutive_model} using time-discrete samples. 
Both the time-discrete and time-series samples are collected from the same microscopic training set. 

\begin{figure}[tbp]
\centering
\includegraphics[width=0.78\textwidth]{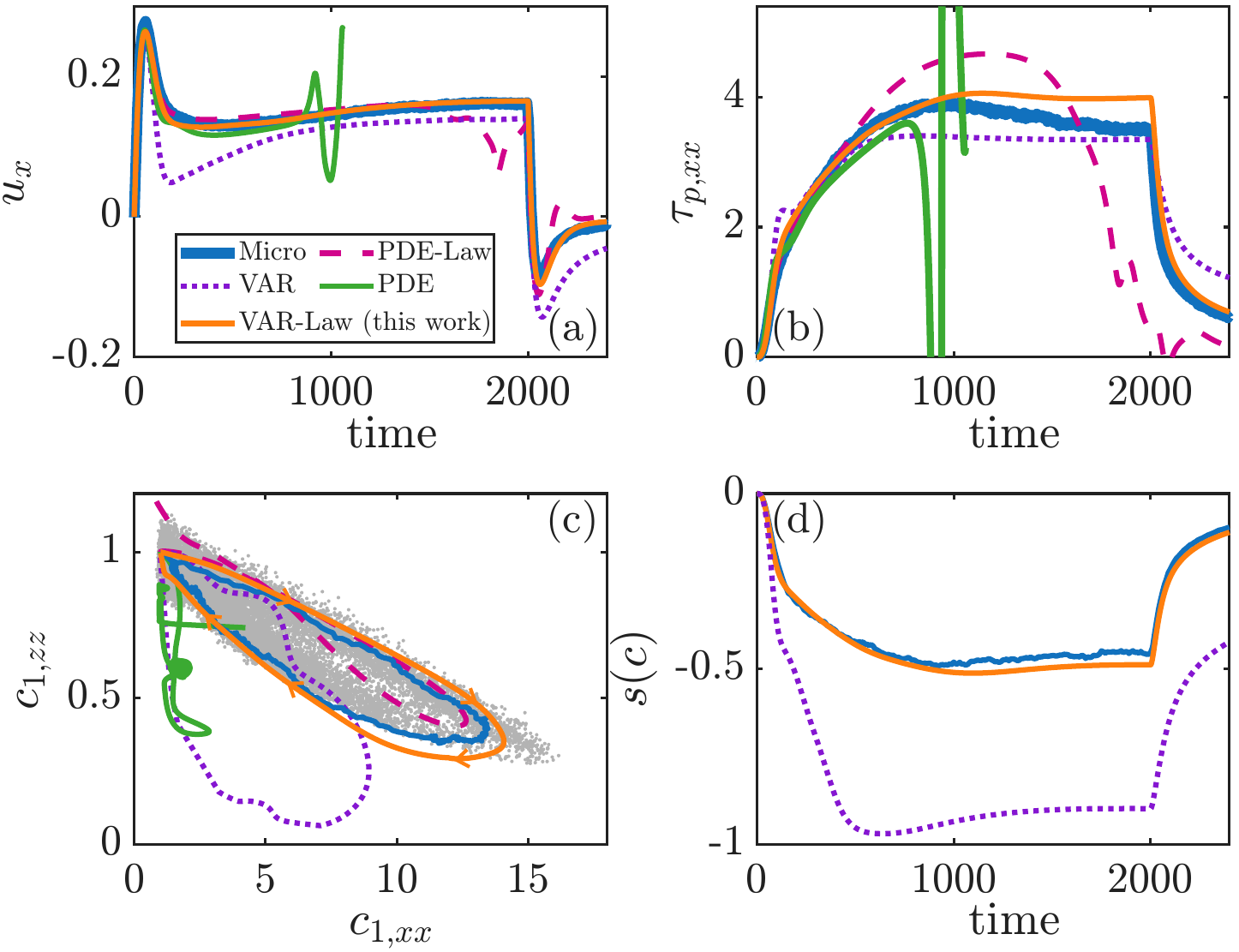}
\caption{Long-time reverse Poiseuille flow of the chain-shaped polymer solution predicted by the micro-model and different macro-models. ``PDE'' and ``PDE-Law'' denote the macro-model in PDE form in Eq.~\eqref{eq:empirical_stress_dynamics} constructed by directly fitting the time-series samples and the explicit micro-model counterparts in Eq. \eqref{eq:micro_consitutive_model}, respectively. ``VAR'' denotes the macro-model in variational form in Eq.~\eqref{eq:variational_PDE_form} by fitting the time-series samples. ``VAR-Law'' denotes the present model. (a) The velocity $u_x(y,t)$ at $y=22.5$. (b) The polymer stress $\tau_{p,xx}(y,t)$ at $y=2.5$. (c) Trajectories of $(c_{1,xx},c_{1,zz})$ at $y=2.5$; the gray points represent the sample points in the training set. (d) Entropy density at $y=2.5$. }
\label{fig:rpf_long}
\end{figure}

Fig. \ref{fig:rpf_long} (a-b) shows the evolution of the velocity and polymer stress predicted from the full microscopic model and the four macroscopic models. The present ``VAR-Law'' model remains stable and agrees well with the full model results. In contrast, the ``PDE'' and ``PDE-Law'' models become unstable at an early and intermediate stage. Accordingly, the conformation tensor trajectories drift out of the training regimes as shown in Fig. \ref{fig:rpf_long}(c). Notably, the ``VAR'' model remains stable due to the preservation of the variational structure. However, the model relies on training using time-derivative samples from the micro-model and yields inaccurate predictions of the conformation tensor and the entropy, as shown in Fig. \ref{fig:rpf_long}(c-d).
By exploiting the analytical microscopic laws and preserving the variational structure, the present model yields both robust and reliable predictions throughout the prolonged simulation.

\subsection{Preservation of physical admissibility} Besides the model robustness, we further examine the physical admissibility of the present model. We simulate the elongation flow of the chain-shaped polymer solution with $\bm\kappa = {\rm diag}(\dot{\epsilon}, -\dot{\epsilon}, 0)$, where $\dot{\epsilon} = 5\times 10^{-4}$. At $t = 4\times 10^4$, we set $\bm\kappa = \bm 0$ and examine the relaxation towards equilibrium. 

\begin{figure}[tbp]
\centering
\includegraphics[width=0.78\textwidth]{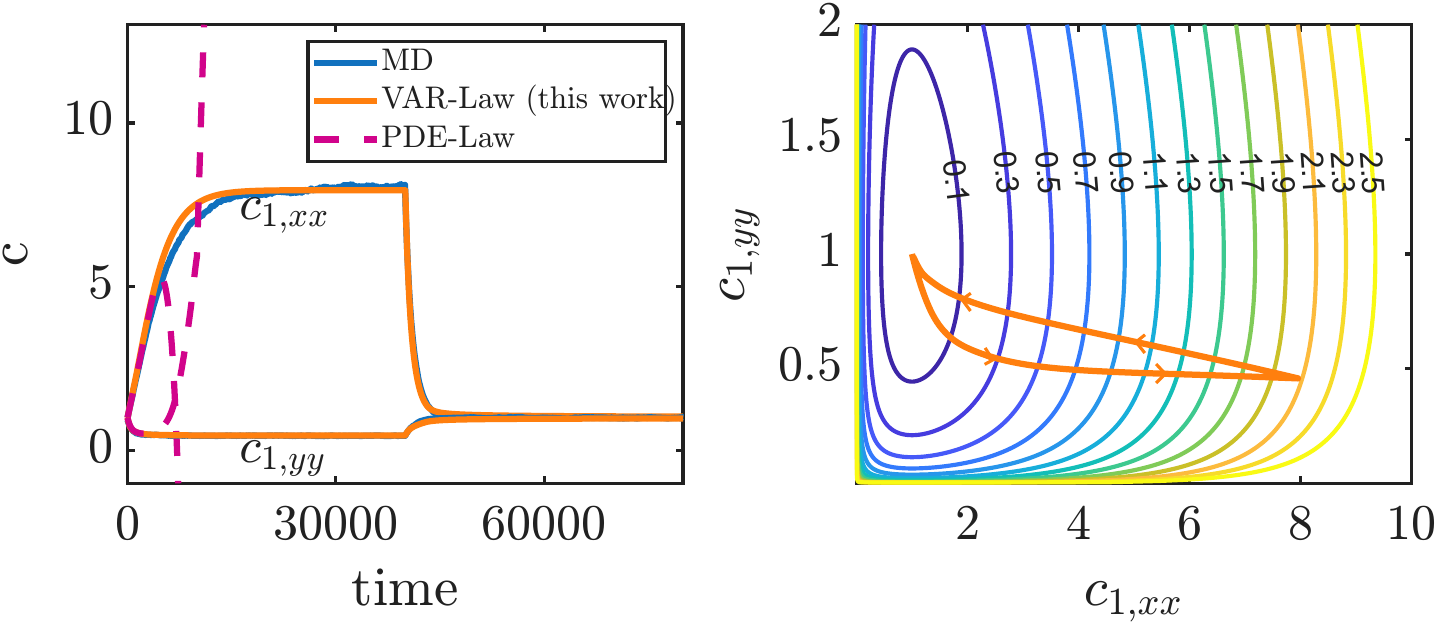}
\caption{The elongation flow of the chain-shaped polymer solution with flow gradient $\bm\kappa =\operatorname{diag}(5\times10^{-4},-5\times10^{-4},0)$ for $0\leq t<4\times10^4$ and is set to zero thereafter. (a) Time evolution of $c_{1,xx}$ and $c_{1,yy}$ predicted by the macroscopic PDE-Law and VAR-Law (present model) as well as the full microscopic model. The prediction from the PDE-Law model becomes unstable and loses positive definiteness near $t=10^4$. (b) Thermodynamic-potential landscape together with the trajectory of the
present model during the elongation and relaxation stages.
The structured entropy term imposes an energy barrier at the boundary of the positive-definite cone and constrains the conformation tensors to be physically admissible.}
\label{fig:elongation_structure}
\end{figure}

Figure~\ref{fig:elongation_structure}(a) shows the evolution of the conformation tensor $\bm c_1$. The prediction from the present model remains positive definite and shows good agreement with the full micro-model results. In contrast, $\bm c_1$ predicted from the PDE-Law model loses positive definiteness during the large deformation at $t = 1\times 10^4$ with similar results for $\bm c_2$ and $\bm c_3$. The preservation of the physical admissibility can be understood by examining the thermodynamic landscape associated with the learned entropy density shown in Figure~\ref{fig:elongation_structure}(b). As the conformation tensor departs from the equilibrium $\bm c_i = \bm I$, the entropy term $\log \det \bm c_i$ imposes a free-energy barrier near the boundary of the positive-definite cone and therefore prevents $\bm c_i$ from approaching singular values. Moreover, the concave structure of the entropy density drives the system back to the unique equilibrium with $\bm \kappa = \bm 0$. 

\subsection{Variationally informed stochastic extension} 
Finally, we consider the stochastic model in Eq. \eqref{eq:stochastic_model} for the dynamics of the conformation tensors under thermal fluctuations. Specifically, we examine the coil--stretch transition of chain polymers under elongation flow, which is a canonical nonequilibrium phenomenon observed in single-molecule experiments \cite{Perkins_Chu_Single_Science_1997}. Here, we use a free-draining bead--spring model as the full microscopic reference similar to Ref. \cite{jendrejack_stochastic_2002,tang_coilstretch_2010}. Due to the competition between the flow-induced stretching and intramolecular potential, the polymer configuration exhibits a transition from the coiled to the extended state, where the transition rate further depends on the bond potential and molecule size. Near the transition regime, the molecule configuration exhibits large fluctuations reflected in the pronounced statistical variance of molecule structures. 

\begin{figure}[tbp]
\centering
\includegraphics[width=0.70\textwidth]{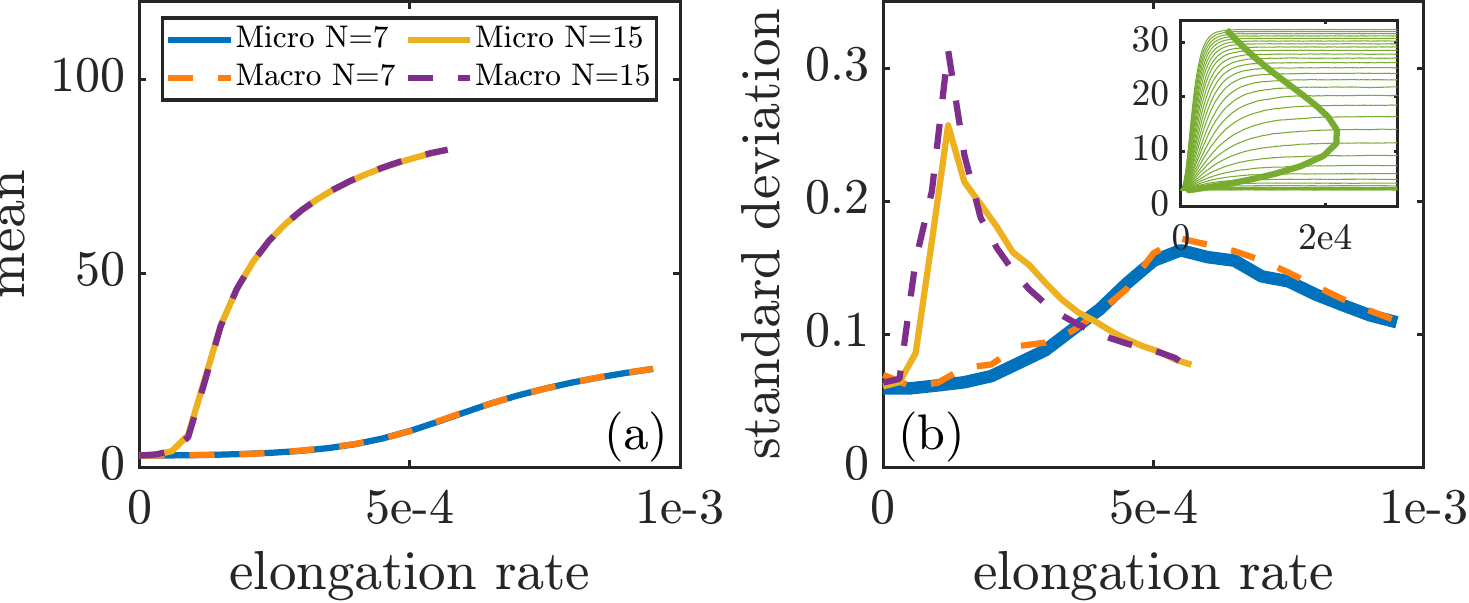}
\caption{Thermal-induced stochastic conformation dynamics of chain-shaped molecules with $N=7$ and $N=15$ under elongational flow over a range of elongation rates $\dot{\epsilon}$. (a) Mean of ${\rm Tr}(\bm c_1)$ (b) Standard deviation of ${\rm Tr}(\bm c_1)$ sampled over $1600$ molecules. The fluctuation peak represents the coil-stretch transition regime where the flow-induced stretching competes with molecular relaxation. The stochastic macro-model accurately captures the cross-over response and its dependence on chain length. 
The inset plot shows the time evolution of ${\rm Tr}(\bm c_1)$ at different elongation rates for the molecule with $N=7$. The thick curve marks the required time when ${\rm Tr}(\bm c_1)$ reaches steady state for each elongation rate. The longest required time near $2.1\times 10^4$ indicates a coil--stretch transition under elongation rate $5\times 10^{-4}$.}
\label{fig:elongation_chain_length}
\end{figure}

Fig. \ref{fig:elongation_chain_length} shows the mean and standard deviation of ${\rm Tr}(\bm c_1)$ for two molecule sizes $N = 7$ and $N = 15$ sampled from the stochastic macro-model in Eq.~\eqref{eq:stochastic_model} and the full micro-model. The transition occurs at a smaller elongation rate with a more pronounced peak for $N = 15$ since the larger molecule has a longer relaxation time and the more pronounced fluctuation peak is due to the size-dependent conformational landscape; see also Ref. \cite{jendrejack_stochastic_2002,tang_coilstretch_2010} for semi-analytical discussions. For both cases, the stochastic macro-model accurately reproduces both the mean conformation response and the standard deviation under fluctuations. This result demonstrates that the present learned variational structure not only describes deterministic constitutive dynamics but also provides a consistent fluctuation--dissipation stochastic closure for non-equilibrium processes.

\section{Summary} This work presents a pathway for transferring analytical microscopic physical laws to reliable macroscopic models. Unlike common approaches that directly learn macroscopic PDEs for prescribed field variables from trajectory data, the present method seeks an autonomous macroscopic closure within an extendable variational structure augmented with auxiliary variables. These variables are represented by symmetry-preserving micro--macro encoders, which establish the explicit microscopic correspondences for their evolution terms. By directly exploiting the analytical microscopic laws, the encoders and the variational ingredients are jointly learned from time-discrete microscopic samples. The resulting macroscopic PDEs inherit the prescribed variational structure, which ensures the free energy stability and preserves the imposed physical constraints by construction. Moreover, this structure provides a basis for the structure-preserving numerical discretizations and a stochastic extension with a consistent fluctuation--dissipation relation.

We demonstrate the present method by learning a macroscopic non-Newtonian hydrodynamic model of dilute polymer solutions directly from a microscopic molecular description. The resulting model naturally encodes the nonlinear intramolecular interactions and the molecular structures, and accurately predicts the viscoelastic responses, where the empirical closures or trajectory-based learning show limitations. Meanwhile, the embedded variational structure yields robust long-time simulations and preserves positive-definite conformation tensors, whereas directly learned PDE models can become unstable. Moreover, the variationally informed stochastic form accurately predicts the coil-stretch transition under elongation flow, where the fluctuation of the polymer configuration exhibits a peak near the transition regime. These results demonstrate both the molecular-level fidelity achieved through microscopic law-level transfer and the numerical robustness of the constructed model. 

While the present work focuses on polymeric fluids, the framework is general and can be extended to multiscale fluids with complex microscopic descriptions such as the hydrodynamic interactions underlying polymer coil--stretch hysteresis \cite{Schroeder_observation_2003}, orientational interactions in nematic fluids \cite{Liu_Sun_DCDS_2009,Forest_Zhou_PRL_2004} and charge interactions in ionic fluids \cite{Eisenberg_Liu_JCP_2010}. More broadly, the joint learning of auxiliary variables and variational structures may provide a new approach to construct reliable kinetic-to-hydrodynamic closures \cite{Han_Ma_PNAS_2019}. We leave these for future studies.

\section*{Data availability}
Micro-model parameters used in the study are included in Appendix~\ref{app:micro_model}, and the training code is available at \url{https://github.com/qdgp/Variational-learning-of-energy-stable-macroscopic-flow-equations-from-microscopic-laws}.

\section*{Author contributions}
PG and HL designed research, performed research, analyzed data, and wrote the paper.

\section*{Conflict of interest}
The authors declare no conflict of interest.

\section*{Acknowledgments}
This work is supported in part by the National Science Foundation under grant DMS-2143739 and the ACCESS program through allocation MTH210005. The authors also acknowledge the support provided by the Institute for Cyber-Enabled Research at Michigan State University.

\begin{appendices}
\numberwithin{equation}{section}
\renewcommand{\theequation}{\thesection\arabic{equation}}
\renewcommand{\thesubsection}{\thesection\arabic{subsection}}
\section{Derivation of the variational hydrodynamic equations for an open system}\label{app:open_system}

Let $\bm\chi=(\bm M,\bm c_1,\ldots,\bm c_{N_c})$, where $\bm M=\rho\bm u$ is the momentum and $\bm c_i\in\mathbb S_+^3$ are generalized conformation tensors. For incompressible flow, $\nabla\cdot\bm M=0$, and variations with respect to $\bm M$ are restricted to divergence-free fields. The evolution of an observable $F[\bm\chi]$ is
\begin{equation}
\dot{F}= \{F, \mathcal{H}\} + (F,\mathcal{H})_{-},
\label{app_eq:general_bracket_variation}
\end{equation}
where the Poisson bracket generates reversible transport and $(F,G)_{-}=(F,G)_{-}^s+(F,G)_{-}^a$ combines a non-positive symmetric part with a skew-symmetric coupling. 

The free energy functional takes the form 
\begin{equation}
\mathcal{H}[\bm\chi] = \int \left [\frac{\vert \bm M \vert^2}{2\rho} - s(\bm c_1, \cdots, \bm c_{N_c}) \right]\diff \bm x,
\label{app_eq:free_energy}
\end{equation}
where $s$ is the macroscopic entropy density and $-s$ is the configuration free energy density. The detailed form will be learned from the microscopic model.  

Given two general observables $F[\bm \chi]$ and $G[\bm \chi]$, the Poisson bracket takes the extendable form
\begin{equation}
\begin{split}
\{F,G\} &=  \int \bm M\cdot
\left[
(G_{M}\cdot\nabla)F_{M} - 
(F_{M}\cdot\nabla)G_{M} 
\right]  \diff \bm x 
 + 
 \sum_{i = 1}^{N_c} \int 
\left[F_i: \mathcal{T}_{G_{M}}\bm c_i  - 
G_i: {\mathcal{T}}_{F_{M}}\bm c_i \right] \diff \bm x,
\end{split}
\label{app_eq:extendable_poisson_bracket}
\end{equation}
where $F_M := \delta F/\delta \bm M$ under constraint $\nabla \cdot \bm M = 0$, $F_{i} := \delta F/\delta \bm c_{i}$ and $\mathcal{T}_{\bm v} \bm c:= -\bm v\cdot\nabla\bm c +(\nabla\bm v)\bm c+\bm c(\nabla\bm v)^\top$.  

The dissipative bracket consists of both symmetric and skew-symmetric parts and takes the form 
\begin{equation}
\begin{split}
(F, G)_{-} &= (F, G)_{-}^s + (F, G)_{-}^a \\ 
(F, G)_{-}^s &= - \int \frac{\eta}{2}  \l({\bm D} F_{M}\r): \l({\bm D} G_{M}\r) \diff \bm x    
-\sum_{i,j=1}^{N_c}  \int_{\Omega} 
F_{i}:\mathcal{M}_{ij}:G_{j}
  \diff \bm x \\
(F, G)_{-}^a &= \sum_{i=1}^{N_c} \int
\left[ F_{i} : (\nabla G_{M}:\mathcal{E}_{i}) 
 -  G_{i} : (\nabla F_{M}:\mathcal{E}_{i}) \right] \diff \bm x,
\end{split}
\label{app_eq:dissipative_operator}
\end{equation}
where $\bm D \bm v := \nabla \bm v + \nabla {\bm v}^{\top}$, $\mathcal{M}_{ij}(\bm c_1, \cdots, \bm c_{N_c}) \in \mathbb{R}^{3\times3\times3\times3}$ and $\mathcal{E}_{i}(\bm c_1, \cdots, \bm c_{N_c}) \in \mathbb{R}^{3\times3\times3\times3}$. The block operator 
$\bm {\mathcal{M}} = [\mathcal{M}]_{i,j=1}^{N_c} \succeq 0$ is self-adjoint, frame-indifferent, and represents the conformation-dependent friction matrices coupling among the individual features. On the other hand, 
$\mathcal{E}_{i}$ represents the additional nonlinear coupling between the flow field $\bm u$ and the individual molecular features $\bm c_{i}$. While it is often set to zero for standard constitutive dynamics, the non-zero term enables us to construct a generalized objective tensor derivative $\mathcal{D} \bm c_{i}/\mathcal{D} t$ for clear microscopic interpretations.

To derive the PDE system, we choose $F$ to take the component-wise form of $\bm\chi(\bm x, t)$ in Eq. \eqref{app_eq:general_bracket_variation}. With $\delta \mathcal{H}/\delta \bm M = \bm u$, we have
\begin{equation}
\begin{aligned}
&\{F,\mathcal{H}\} = \int\! \bm M \cdot \left[ (\bm u \cdot\nabla)F_{M} - (F_{M}\cdot\nabla)\bm u \right] \diff \bm x + \sum_{i = 1}^{N_c} \int\! \left[F_i: \mathcal{T}_{\bm u}\bm c_i - \mathcal{H}_i: {\mathcal{T}}_{F_{M}}\bm c_i \right] \diff \bm x,
\\
&= -\int\! M_k \l( \tfrac{\delta F}{\delta M_j} \tfrac{\partial }{\partial x_j} u_k - u_j \tfrac{\partial }{\partial x_j} \tfrac{\delta F}{\delta M_k} \r) \diff \bm x  - \sum_{i=1}^{N_c} \int\! \tfrac{\delta F}{\delta c_{i,jk}} \l( \tfrac{\partial }{\partial x_l} c_{i,jk} \r) u_l + \tfrac{\partial s}{\partial c_{i,jk}} \l( \tfrac{\partial }{\partial x_l} c_{i,jk} \r) \tfrac{\delta F}{\delta M_l} \diff \bm x
\\
&+\sum_{i=1}^{N_c} \int\! c_{i,lk} \l(\tfrac{\delta F}{\delta c_{i,jk}} \tfrac{\partial }{\partial x_l} u_j + \tfrac{\partial s}{\partial c_{i,jk}} \tfrac{\partial }{\partial x_l} \tfrac{\delta F}{\delta M_j} \r) \diff \bm x  +\sum_{i=1}^{N_c} \int\! c_{i,jl} \l(\tfrac{\delta F}{\delta c_{i,jk}} \tfrac{\partial }{\partial x_l} u_k + \tfrac{\partial s}{\partial c_{i,jk}} \tfrac{\partial }{\partial x_l} \tfrac{\delta F}{\delta M_k} \r) \diff \bm x
\end{aligned}
\label{app_eq:H_extendable_poisson_bracket}
\end{equation}
where Einstein index summation is applied to indices $j$, $k$ and $l$.  To proceed, we consider $F_1 = \int f_1(\bm M) \diff \bm x$, and collect the variational terms in Eq. \eqref{app_eq:H_extendable_poisson_bracket} with respect to $\delta F_1/\delta \bm M$ by 
\begin{equation}
\begin{split}
\{F_1,\mathcal{H}\} &=  -\int M_k \left( \fd{F}{M_j} \fp{}{x_j} u_k - u_j \fp{}{x_j} \fd{F}{M_k} \right) \diff \bm x 
- \sum_{i = 1}^{N_c} \int \fp{s}{c_{i, jk}} \l( \fp{}{x_l} c_{i,jk} \r) \fd{F}{M_l} \diff \bm x \\
 &~+\sum_{i = 1}^{N_c} \int c_{i, lk} \l(+ \fp{s}{c_{i,jk}}  \fp{}{x_l} \fd{F}{M_j}  \r) \diff \bm x 
 +\sum_{i = 1}^{N_c} \int c_{i, jl} \l( + \fp{s}{c_{i,jk}}  \fp{}{x_l} \fd{F}{M_k}  \r) \diff \bm x  \\ 
 =  & \int \l[ - M_j \fp{}{x_k} u_j -  \fp{}{x_j} M_k   u_j 
- \fp{s}{x_k} -  \fp{}{x_l} \sum_{i = 1}^{N_c} \left( c_{i,lj} \fp{s}{c_{i,kj}}
+  c_{i,jl} \fp{s}{c_{i,jk}} \right)
\r] \fd{F}{M_k} \diff \bm x,
\end{split}    
\end{equation}
which gives 
\begin{equation}
\rho \dot{u}_k = \mathbb{P} \left[- M_j \nabla_k u_j - M_j \nabla_j u_k -  \fp{s}{x_k}  -  \sum_{i = 1}^{N_c} 2\nabla_l  \left(c_{i, lj} \fp{s}{{c}_{i, jk}}\right)\right],     
\end{equation}
where $\mathbb{P}$ represents the Leray--Hodge projection onto the divergence-free space, i.e., $\mathbb{P} \bm v = \bm v - \nabla \Delta^{-1} (\nabla \cdot \bm v)$.  By introducing the pressure term $p$ as the source term of the divergence and absorbing the $\nabla s$ term, we have
\begin{equation}
\begin{split}
\nabla \cdot \bm u &= 0 \\
\rho \dot{\bm {u}}  &= -\rho \bm u \cdot \nabla \bm u - \nabla p - \sum_{i = 1}^{N_c} \nabla \cdot \l( 2 \bm c_i \fp{s}{\bm c_i} \r)  
\end{split}
\label{app_eq:hydrodynamics_reversible}
\end{equation}
Similarly, we consider $F_2 = \int f_2(\bm c_i) \diff \bm x$, and collect the variational terms in Eq. \eqref{app_eq:H_extendable_poisson_bracket} with respect to $\delta F/\delta \bm c_i$ by 
\begin{equation}
\begin{split}
\{F_2,\mathcal{H}\} = - \int \tfrac{\delta F}{\delta c_{i,jk}}  \l( \tfrac{\partial }{\partial x_l} c_{i,jk} \r)   u_l   \d^3 x 
+\int c_{i,lk} \l(\tfrac{\delta F}{\delta c_{i,jk}} \tfrac{\partial }{\partial x_l} u_j \r) \diff \bm x 
+\int c_{i,jl} \l(\tfrac{\delta F}{\delta c_{i,jk}} \tfrac{\partial }{\partial x_l} u_k \r) \diff \bm x,
\end{split}    
\end{equation}
which gives 
\begin{equation}
\dot{\bm c_i} = -\bm u \cdot \nabla \bm c_i + \bm \kappa^\top \bm c_i + \bm c_i \bm \kappa. 
\label{app_eq:upper_convective}
\end{equation}
where $\bm\kappa=\nabla\bm u^\top$. Equation~\eqref{app_eq:upper_convective} gives the upper-convected transport. The skew-symmetric dissipative bracket adds the flow--conformation coupling derived below. 

For the dissipative part $(F, \mathcal{H})_{-}$, we have 
\begin{equation}
\begin{split}
(F, \mathcal{H})_{-} &= (F, \mathcal{H})_{-}^s + (F, \mathcal{H})_{-}^a   \\
&= - \int \frac{1}{2} \eta \l(
\tfrac{\partial }{\partial x_k} \tfrac{\delta F}{\delta M_{l}} + \tfrac{\partial }{\partial x_l} \tfrac{\delta F}{\delta M_{k}} \r)\l( \tfrac{\partial }{\partial x_k} u_l + \tfrac{\partial }{\partial x_l} u_k \r)  \diff \bm x
+ \sum_{i,j=1}^{N_c} \int \tfrac{\delta F}{\delta c_{i}} : \mathcal{M}_{ij} : \tfrac{\partial s}{\partial c_{j}}   \diff \bm x \\      
&\quad + \sum_{j=1}^{N_c} \int \mathcal{E}_{j, klpq} \l( \tfrac{\partial s}{\partial c_{j, pq}} \tfrac{\partial }{\partial x_k} \tfrac{\delta F}{\delta M_{l}} + \tfrac{\delta F}{\delta c_{j, pq}} \tfrac{\partial }{\partial x_k} u_l \r) \diff \bm x \\
\end{split}
\end{equation} 
We consider $F_1 = \int f(\bm M) \diff \bm x$ and $F_2 = \int f_2(\bm c_i) \diff \bm x$ and collect the variational terms with respect to $\delta F/\delta \bm M$ and  $\delta F/\delta \bm c_i$, i.e., 
\begin{equation}
\begin{split}
(F_1, \mathcal{H})_{-}  &= \int  \fd{F}{M_{l}} \left [\eta \frac{\partial^2}{\partial x_k^2}   u_l  -  \sum_{j=1}^{N_c}  \fp{}{x_k} \l( \mathcal{E}_{j,klpq}  \fp{s}{c_{j,pq}} \r) \right] \diff \bm x, \\
(F_2, \mathcal{H})_{-}  &=  \sum_{j=1}^{N_c} \int \fd{F}{\bm c_{i}} :\mathcal{M}_{ij} : \fp{s}{\bm c_{j}}  \diff \bm x + \int \mathcal{E}_{i,klpq} \l( \fd{F}{c_{i,pq}} \fp{}{x_k} u_l \r) \diff \bm x \\
&= \sum_{j=1}^{N_c} \int \fd{F}{\bm c_{i}} :\mathcal{M}_{ij} : \fp{s}{\bm c_{j}}  \diff \bm x + \int \bm\kappa : \mathcal{E}_{i} :  \fd{F}{\bm c_{i}} \diff \bm x 
\end{split}
\end{equation}
which gives the irreversible dynamics in the form of 
\begin{equation}
\begin{split}
\rho \dot{\bm u} &= \mathbb{P} \left[ \eta \nabla \cdot \nabla \bm u - \nabla \cdot \l(\sum_{j=1}^{N_c} \mathcal{E}_j: \fp{s}{\bm c_j}\r) \right]  \\
\dot{\bm c_i} &=  \bm \kappa : \mathcal{E}_{i} + \sum_{j=1}^{N_c} \mathcal{M}_{ij} : \frac{\partial s}{\partial \bm c_j}
\end{split}
\label{app_eq:hydrodynamics_irreversible}
\end{equation}

Combining Eqs.~\eqref{app_eq:hydrodynamics_reversible} and \eqref{app_eq:hydrodynamics_irreversible} and adding the external body force $\bm g$ gives the complete open-system equations 
\begin{equation}
\begin{split}
\nabla\cdot\bm u &=0 \\
\rho\left(\partial_t\bm u
+\bm u\cdot\nabla\bm u\right)
&=
-\nabla p+ \eta \Delta \bm u + \nabla\cdot \bm \tau_p + \bm g\\
\partial_t \bm {c}_{i}
+ \bm u \cdot \nabla \bm {c}_{i} - \bm \kappa : \tilde{\mathcal{E}}_{i} & =  \sum_{j=1}^{N_c} \mathcal{M}_{ij} :\frac{\partial s}{\partial \bm{c}_{j}} \\
\bm\tau_p &= - \sum_{i=1}^{N_c} \tilde{\mathcal{E}}_{i}: \frac{\partial s}{\partial \bm c_{i}}, 
\end{split}
\label{app_eq:variational_PDE_form}
\end{equation}
where $\bm \kappa = \nabla \bm u^{\top}$, $(\widetilde{\mathcal E}_i)_{klmn}
= (\mathcal E_i)_{klmn} + (c_i)_{km}\delta_{ln}
+(c_i)_{kn}\delta_{lm}$.  

The corresponding free-energy balance is
\begin{equation}
\frac{\diff \mathcal H}{\diff t} = (\mathcal{H}, \mathcal{H})_{-} + \int \bm g \cdot \bm u \diff x    
\end{equation}
which reduces to $\dot{\mathcal{H}} \le 0$ for $\bm g \equiv 0$. 

\section{Extendable variational form of hydrodynamic equations for closed systems}\label{app:closed_system}
While this work uses an open system to illustrate the essential idea, the present method can be used for modeling closed systems. Following the GENERIC framework \cite{Ottinger_Grmela_1997_2, Ottinger_beyond_2005}, we can further split the free energy $\mathcal{H}$ into the energy $E$ and the entropy $S$ and construct the extendable variational form that preserves energy conservation and entropy production. To construct the macroscopic model, we represent the time evolution of a general observable $F[\chi]$ in the variational form 
\begin{equation}
\dot{F}= \{F, E\} - (F, S)_{-},
\label{app_eq:close_general_bracket_variation}
\end{equation}
where $\{\cdot, \cdot\}$ is the skew-symmetric Poisson bracket and $(\cdot, \cdot)_{-}$ is the dissipative bracket similar to the open system. Crucially, the two brackets satisfy the following degeneracy conditions
\begin{equation}
\{F, S\} \equiv 0 \quad (F, E)_{-} \equiv 0, 
\label{app_eq:closed_degeneracy_condition}
\end{equation}
which implies that $S$ is a Casimir invariant and $E$ is conserved in the dissipation process. Therefore, Eq. \eqref{app_eq:close_general_bracket_variation} preserves the energy conservation $\dot{E} = 0$ and non-negative entropy production $\dot{S} \ge 0 $.

Here, we briefly discuss the variational form for a closed incompressible non-isothermal non-Newtonian fluid system. For simplicity, we assume the polymer number density is a constant. 
Let $\bm\chi = (\bm M, e, \bm c_1, \cdots, \bm c_{N_c})$ denote a set of macroscopic field variables, where $\bm M = \rho \bm u$ is the momentum under the divergence-free constraint $\nabla \cdot \bm M = 0$ and $e(\bm x, t)$ denotes the internal energy. 
The energy $E$ and entropy $S$ takes the form 
\begin{equation}
E[\bm \chi] = \int  \left [\frac{\vert \bm M \vert^2}{2\rho} + e \right]\diff \bm x, \quad S[\bm \chi] = \int  \left [s(e, \bm c_1, \cdots, \bm c_{N_c}) \right]\diff \bm x,
\label{app_eq:energy_entropy}
\end{equation}
where $s$ is the entropy density. Similar to the notation for the open system, we use $F_e$ to denote $\delta F/\delta e$. Accordingly, $S_e := \partial s/\partial e = 1/T$ is the inverse of temperature $T$. 

Given two general observables $F[\bm \chi]$ and $G[\bm \chi]$, we construct the extendable form of the Poisson bracket 
\begin{equation}
\begin{aligned}
\{F,G\}&={}\int\bm M\cdot \left[(G_{\bm M}\cdot\nabla)F_{\bm M} -(F_{\bm M}\cdot\nabla)G_{\bm M}\right]\diff \bm x
\\
&+\sum_i\int \left[F_i:\mathcal T_{G_{\bm M}}\bm c_i -G_i:\mathcal T_{F_{\bm M}}\bm c_i\right] \diff \bm x +\int \left[F_e\mathcal Q_{G_{\bm M}} -G_e\mathcal Q_{F_{\bm M}}\right]\diff \bm x
\end{aligned}
\label{app_eq:closed_extendable_poisson_bracket}
\end{equation}
where $\mathcal Q_{\bm  v} :=-\bm  v\cdot\nabla e
-\boldsymbol\pi:\nabla\bm  v$ for $\bm v \in \mathbb{R}^3$ and $\bm \pi$ represents the reversible part of the stress tensor whose detailed form will be determined by the degeneracy condition i.e., 
$\{S,F\} =-\int \bm  v\cdot\nabla s\, \diff \bm  x
+\int \left[
\sum_i \fp{s}{\bm{c}_i}:
\big((\nabla\bm  v)\bm  c_i+
\bm  c_i(\nabla\bm  v)^\top\big)
-\frac{\boldsymbol\pi}{T}:\nabla\bm  v
\right] \diff \bm  x = 0$,
which gives $
\boldsymbol\pi
=T\sum_{i=1}^{N_c} 2
\bm c_i \frac{\partial s}{\partial \bm c_i}$.

The dissipative bracket takes the extendable form 
\begin{equation}
\begin{split}
(F, G)_{-} &= (F, G)_{-}^s + (F, G)_{-}^a \\ 
(F, G)_{-}^s &= -\int\left[
\frac{\eta T}{2}\mathcal D_F:\mathcal D_G
+T\sum_{i,j}F_i:\mathcal M_{ij}:G_j
+\lambda T^2\nabla F_e\cdot\nabla G_e
\right] \diff \bm  x \\
(F, G)_{-}^a & =\sum_i\int T
\left[ 
F_i:(\mathcal K_G:\mathcal E_i)
-G_i:(\mathcal K_F:\mathcal E_i)
\right] \diff \bm  x ,
\end{split}
\label{app_eq:closed_extendable_dissipative_operator}
\end{equation}
where $ \mathcal D_F := \bm D(F_{\bm  M})- F_e \bm D(\bm  u)$, $\mathcal K_F = (\nabla F_{\bm  M})^\top - F_e \nabla \bm  u^\top$, and $\lambda$ represents the heat conduction coefficient. We can show that the degeneracy condition $(F, E)_{-}^s = (F, E)_{-}^a = 0$ holds.

By using Eqs. \eqref{app_eq:energy_entropy}, \eqref{app_eq:closed_extendable_poisson_bracket}, and \eqref{app_eq:closed_extendable_dissipative_operator}, we have the PDE form for the closed system 
\begin{equation}
\begin{split}
\nabla\cdot\bm  u&=0\\
\rho_0(\partial_t\bm  u+\bm  u\cdot\nabla\bm  u)
&=-\nabla p+ \eta \Delta \bm u + \nabla\cdot\boldsymbol\tau_p\\
\partial_t\bm  c_i+\bm  u\cdot\nabla\bm  c_i
-\kappa:\widetilde{\mathcal E}_i
&=T\sum_j\mathcal M_{ij}:\frac{\partial s}{\partial \bm c_j} \\
\partial_t e+\bm  u\cdot\nabla e
&=\boldsymbol\tau_p: \bm \kappa
+\frac{\eta}{2} \bm D\bm  u: \bm D\bm  u
+\nabla\cdot(\lambda\nabla T)\\
\boldsymbol\tau_p
&=-T\sum_i\widetilde{\mathcal E}_i:\frac{\partial s}{\partial \bm c_i}.   
\end{split}    
\label{app_eq:closed_variational_hydrodynamic_PDE}
\end{equation}
With the variational structure and degeneracy condition by construction, we can show Eq. \eqref{app_eq:closed_variational_hydrodynamic_PDE} satisfies $\dot{E} = 0$ and $\dot{S} \ge 0$. 

While we consider incompressible fluids, the present form is general and can be further adapted for other multiscale systems such as compressible non-Newtonian hydrodynamics, nematic fluids, and kinetic-to-hydrodynamic closures.  

\section{Structure-preserving neural-network representation}\label{app:nn_representation} 
The extendable variational structure is represented by symmetry-preserving NNs for $\left\{\bm c_i\right\}_{i=1}^{N_c}$, the entropy density $s(\bm c_1, \cdots, \bm c_{N_c})$, friction operator $\bm{\mathcal{M}}(\bm c_1, \cdots, \bm c_{N_c})$ and the coupling tensor $\mathcal{E}(\bm c_1, \cdots, \bm c_{N_c})$. Below we show their detailed forms that preserve the physical constraints and frame-indifference symmetries, as well as the detailed NN parameters.  

\subsection{Micro-macro encoders} The macroscopic auxiliary variables $\left\{\bm c_i\right\}_{i=1}^{N_c}$ are constructed by seeking a set of micro-macro encoder functions $\{\bm b_i(\bm r)\}_{i=1}^{N_c}$ that map the microscopic polymer molecule configurations $\bm r$ to a set of macroscopic second-order tensor-valued features, i.e., 
\begin{equation}
   \bm c_i = \left\langle \bm b_i(\bm r)\right\rangle \quad \bm b_i = \bm h_i \bm h_i^\top \quad \bm h_i = g_i(\bm q^{\ast}) \sum_{k=1}^{N-1} w_{ik} \bm r_k, \label{app_eq:encoder_b}
\end{equation} 
where $\bm q^{\ast} \in \mathbb{R}^{3N-6}$ is a translational- and rotational-invariant representation of the molecular configuration. One straightforward choice is 
\begin{equation*}
\bm q^{\ast} = \left[
\left| \bm{r}_{1} \right|,
\left| \bm{r}_{2} \right|,
\left| \bm{r}_{12} \right|,
\left| \bm{r}_{3} \right|,
\left| \bm{r}_{13} \right|,
\left| \bm{r}_{23} \right|,
\cdots,
\left| \bm{r} _{N-1} \right|,
| \bm{r} _{(N-2)(N-1)} |
\right]
\end{equation*}
 and $\{\bm r_k\}_{k=1}^{N-1}$ is a set of $N-1$ independent bond vectors selected from $\bm r$. 
$g:\mathbb{R}^{3N-6}\to\mathbb{R}^{N_c}$ is a NN and $\bm w \in \mathbb{R}^{N_c\times (N-1)}$ are the weights. In practice, $g_i$ is rescaled using a normalization factor
determined from $\operatorname{Tr}
\langle\bm b_i(\bm r)\rangle_{\rm eq}$,
such that $\bm c_i^{\rm eq}=\bm I$. We can show $\bm b_i$ satisfy the frame-indifferent constraint, i.e., $\bm b_i(\bm r\mathcal{U}^\top) = \mathcal{U} \bm b_i \mathcal{U}^\top$, $\mathcal{U}\in SO(3)$. 

In this study, we represent $g$ using a fully-connected NN with $3$ hidden layers and $100$ neurons per layer. The activate function is hyperbolic tangent. Both  $g$ and $\bm w$ will be learned from the microscopic model. 

\subsection{Entropy density}  The polymer entropy density $s$ is required to be frame-indifferent, concave, and maximized at the prescribed equilibrium state. For brevity,
we denote $\bm c=(\bm c_1,\ldots,\bm c_{N_c})$ and represent
\begin{equation}
\begin{split}
s(\bm c) &= \psi_s(\bm c) - \left .\frac{\partial \psi_s}{\partial \bm c}\right \vert_{\bm c^{\rm eq}} : \bm c \\
\psi_s(\bm c) &= -\psi_e(\bm c) + s_c(\bm c) \quad s_c(\bm c) = \sum_{i=1}^{N_c} \lambda_i \log {\det} \bm c_i,
\end{split}
\label{app_eq:entropy_density}
\end{equation}
where $\lambda_i > 0$. Here, $\psi_e$ represents the learned effective elastic free energy
contribution. We represent
$\psi_e(\bm c)=\Phi_\theta(\bm x_c)$, where
$\bm x_c= \left\{
{\rm Tr}(\bm c_i),\, {\rm Tr}(\bm c_i^2),\, {\rm Tr}(\bm c_i^3), {\rm Tr}\!\left[(\bm c_i-\bm c_j)^2\right]
\right\}_{i,\,i<j}$ and $\Phi_\theta$ takes a modified form of the input-convex neural network \cite{ICNN_Amos_Xu_2017}, i.e., 
\begin{equation}
\begin{split}
z_{1} &= l_0(W_0^x \bm x_c+b_0) \\
z_{i+1} &= l_i(W_i^z z_{i}+W_i^x \bm x_c+b_i) \\
\Phi_\theta &= z_{N_\Phi}    
\end{split}
\end{equation}
where $l_i$ is convex and non-decreasing activate functions. Unlike Ref. \cite{ICNN_Amos_Xu_2017}, we require both $W_i^z$ and $W_i^x$ are non-negative. Since $\bm x_c$ is convex with respect to $\bm c_i$,  $\psi_e(\bm c)$ is convex with respect to $\bm c_i$. Therefore, $s(\bm c)$ is concave, frame-indifferent, and achieves the maximum value at $\bm c^{\rm eq}$.

In this study, we represent $\Phi_\theta(\bm x_c)$ using a fully-connected NN with $2$ hidden layers with $100$ neurons per layer. The activate function is softplus.

\subsection{Friction operator and coupling tensor} The friction operator $\bm{\mathcal{M}} (\bm c_1,\ldots,\bm c_{N_c})$ and $\mathcal{E}(\bm c_1,\ldots,\bm c_{N_c})$ are fourth-order tensor functions which need to be frame-indifferent and non-negative ($\bm {\mathcal{M}} \succeq 0$). Let $\mathcal K$ denote $\mathcal E_i$ or $\mathcal M_{ij}$, we require 
\begin{equation}
\begin{split}
\widetilde{\bm R:\mathcal K(\bm c)}
&=
\widetilde{\bm R}:\mathcal K(\widetilde{\bm c})
\qquad
\widetilde{\mathcal K(\bm c):\bm R}
=
\mathcal K(\widetilde{\bm c}):\widetilde{\bm R}
\\
[\bm R:\mathcal K(\bm c)]^\top &= \bm R:\mathcal K(\bm c)
\quad
[\mathcal K(\bm c):\bm R]^\top = \mathcal K(\bm c):\bm R,
\end{split}
\label{app_eq:fourth_order_tensor_symmetry}
\end{equation}
where $\bm R\in\mathbb R^{3\times 3}$, $\widetilde{\bm R}=\mathcal U\bm R\mathcal U^\top$,
$\widetilde{\bm c}=\mathcal U\bm c\mathcal U^\top$ and
$\mathcal U\in SO(3)$. The second line of Eq. \eqref{app_eq:fourth_order_tensor_symmetry} is due to the constraint that the terms in Eq. \eqref{app_eq:variational_PDE_form} such as $\bm\kappa: \mathcal{E}_i$ and $\mathcal{M}_{ij} \frac{\partial s}{\partial \bm c_i}$ need to be symmetric. 

To impose these constraints, we introduce a set of symmetric, rotationally equivariant polynomial tensor basis functions $\{\bm p_s(\bm c)\}_{s=1}^{N_p} =  \{ \bm {I}, \bm c_i, \bm  c_i^2, \bm  c_i^3, \bm  c_i \bm  c_j + \bm  c_j \bm c_i\}_{i,j = 1}^{N_c}$. Pairwise products of these tensors are
combined into the fourth-order blocks designed to enforce
Eq.~\eqref{app_eq:fourth_order_tensor_symmetry} by representing
$\mathcal E$ and $\mathcal M$ as 
\begin{equation}
\begin{split}
& \{\bm {P}_r\}_{r=1}^{N_r}
= \left\{ \bm p_s\boxtimes\bm p_t
+ \bm p_t\boxtimes\bm p_s, 
\bm p_s\otimes\bm p_t 
\right\}_{s,t=1}^{N_p}\\
&\mathcal E_i 
=
\sum_r \xi_{ri}\bm {P}_r, \quad 
\Gamma_{ij}
=
\sum_r \zeta_{r,ij} \bm {P}_r, \quad 
\boldsymbol{\mathcal M} = \boldsymbol\Gamma\boldsymbol\Gamma^\ast + \epsilon \bm{\mathcal{I}}
\end{split}
\label{app_eq:fourth_order_basis_E_M}
\end{equation}
where $(\bm p\boxtimes \bm q)_{klmn}= p_{km}q_{ln} + q_{kn}p_{lm} $ and $(\bm p\otimes \bm q)_{klmn}= p_{kl}q_{mn}$.  $\bm \xi_{r}(\bm z_c) \in \mathbb{R}^{N_c}$ and $\bm \zeta_r (\bm z_c) \in \mathbb{R}^{N_c \times N_c} $ are NNs of rotational
invariants $\bm z_c \in \{ \Tr(\bm p_s(\bm c)) \}_{s=1}^{N_p}$. $\bm\Gamma^\ast$ denotes the
adjoint of $\bm\Gamma$ and $\{\boldsymbol\Gamma\boldsymbol\Gamma^\ast \}_{ij,klmn} = \sum_{w,pq} \boldsymbol\Gamma_{iw,klpq} \boldsymbol\Gamma_{jw,mnpq}$. The term $\epsilon \bm{\mathcal{I}}$ with $\epsilon > 0$ and $\mathcal{I}_{ij} = \delta_{ij}(\delta_{km}\delta_{ln} + \delta_{kn}\delta_{lm})$ is a regularization term for $\bm{\mathcal{M}} \succ 0$. 

In this study, we represent $\bm \xi_{r}(\bm z_c)$ and $\bm \zeta_r (\bm z_c)$ using a fully-connected NN with $2$ hidden layers with $100$ neurons per layer and the activate function is hyperbolic tangent.

\section{Microscopic model description}\label{app:micro_model}
In this study, we represent the microscopic model by the suspension of the bead-spring polymer model. Both the polymer beads and the solvent particles are modeled by the dissipative particle dynamics (DPD) \cite{Hoogerbrugge_SMH_1992, Espanol_SMO_1995, Groot_DPD_1997}. The polymer bond interaction takes the form of the FENE potential  
\begin{equation}
V(\bm r)=\sum_{n=1}^{N_b}V_b\!\left(|\bm r_n|\right), \quad 
V_b (l) =-\frac{k_s}{2}l^2_{0} \log\left[ 1-\frac{l^2}{l^2_{0}}\right],
\label{app_eq:micro-scale_model}
\end{equation}
where $\bm r = \bm K \bm{q}$ is the bond vector, $\bm{q} \in \mathbb{R}^{N\times 3}$ is bead position, and $\bm K\in\mathbb{R}^{N_b\times N}$ is the connectivity matrix determined by the molecule structure. $k_s$ represents the spring constant, and $l_0$ is the maximum extension length. In particular, $V_b$ is nonlinear and diverges when $l$ approaches $l_0$, and therefore, there exists no analytical macroscopic closed form for this microscopic description. Existing closures such as FENE-P can be derived based on the mean-field approximation. 

For the reverse Poiseuille flow, we consider the suspensions of both the chain- and star-shaped molecules. The simulation domain is $60\times100\times60$ with periodic boundary conditions imposed along the three directions. Both types of molecules consist of $N = 7$ beads and $N_b = 6$ bonds with the same bond parameters $k_s = 0.1$ and $l_0 = 2.3$ (in reduced units). The number density is $n_p = 0.3$. The star-shaped molecule consists of three branches. Each branch consists of three beads, with one bead shared by the three branches. For the elongation flow, we consider the suspension of the chain-shaped molecule with $N = 7$, $n_p = 0.3$ and $N = 15$, $n_p = 0.14$; $k_s$ and $l_0$ remain the same as in the reverse Poiseuille flow. The simulation domain is $200\times200\times60$ with the generalized uniaxial extension flow boundary condition  \cite{Nicholson_UEF_JCP_2016, Murashima_Hagita_UEFEX_2018}.

The solvent is modeled by DPD particles with number density $n_s = 4.0$.
The interactions between the particles $i$ and $j$ takes the standard pairwise form
\begin{equation}
\begin{split}
&\mathbf{F}_{ij} = \mathbf{F}_{ij}^C +  \mathbf{F}_{ij}^D +   \mathbf{F}_{ij}^R, \quad \quad \quad\quad \quad \quad \quad\quad\quad~
\mathbf{F}_{ij}^C = 
\begin{cases}
a (1.0 - r_{ij}/r_c) \mathbf{e}_{ij}, & r_{ij} < r_c \\
    0, &r_{ij} > r_c
\end{cases}, 
\\
&\mathbf{F}_{ij}^D = 
\begin{cases}
-\gamma w^{D}(r_{ij})(\mathbf{v}_{ij} \cdot \mathbf{e}_{ij}) \mathbf{e}_{ij}, & r_{ij} < r_c\\
    0, &r_{ij} > r_c
\end{cases},  \quad   
\mathbf{F}_{ij}^R = 
\begin{cases}
\sigma w^{R}(r_{ij}) \xi_{ij} \mathbf{e}_{ij}, & r_{ij} < r_c\\
    0, &r_{ij} > r_c
\end{cases},
\end{split}\nonumber
\end{equation}
where $\mathbf{r}_{ij} = \mathbf{r}_{i} - \mathbf{r}_{j}$, $r_{ij} = |\mathbf{r}_{ij}|$, $\mathbf{e}_{ij} = 
\mathbf{r}_{ij}/r_{ij}$, and $\mathbf{v}_{ij} = \mathbf{v}_{i} - \mathbf{v}_{j}$,
$\xi_{ij}$ are independent identically distributed Gaussian random variables with zero mean and unit 
variance. $\gamma$ and $\sigma$ are related to the system temperature by the second fluctuation-dissipation 
theorem \cite{Espanol_SMO_1995} as $\sigma^2 = 2 \gamma k_BT$, where $k_BT$ is set to $0.25$. The  
parameters remain the same for different flows and are given in Tab. \ref{tab:polymer_model_parameter}.
\begin{table}[htbp]
\centering
\caption{Parameters (in reduced unit) of the micro-scale model of the polymer solution (S-sovlent, P-polymer). }
\label{tab:polymer_model_parameter}
\begin{tabular}{lrrrrr}
\toprule
&  $a$ & $\gamma$ & $\sigma$ & $k$ & $r_c$\\
\midrule
\text{S-S} & $4.0$ & $5.0$ & $1.58$ &$0.25$ & $1.0$\\ 
\text{S-P} & $0.0$ & $40.0$ & $4.47$ &$0.0$ & $1.0$\\ 
\text{P-P} & $0.002 $ & $0.005$ & $0.05$ &$0.5$ & $0.7$\\
\bottomrule
\end{tabular}
\end{table}

\FloatBarrier
\section{Micro-macro learning and training details}\label{app:training}

In this section, we present the micro-model counterparts and the training details for the present model, denoted by ``VAR-Law'', which passes the microscopic physical laws to the macroscopic variation structures. For comparison, we also construct the macroscopic models based on either directly learning the PDE form and/or fitting the time-series samples, denoted by ``PDE-Law'', ``VAR'', and ``PDE'' model. 

\subsection{Microscopic model correspondence}
By establishing the joint learning of a set of micro-macro encoder $\bm c_i =\left\langle \bm b_i (\bm r) \right \rangle$, we can establish the micro-macro correspondence such that the individual unclosed terms in the macroscopic PDE form in Eq. \eqref{app_eq:variational_PDE_form} has a clear microscopic counterpart with clear physical interpretation.  

In this study, we consider the micro-model as the common overdamped Langevin dynamics with constant friction coefficient $\gamma$. 
The evolution of the polymer configuration distribution $f(\bm r, \bm x, t)$ is governed by the Fokker-Planck equation. Accordingly, we can derive the evolution of the conformation tensor $\bm c_i = \int f(\bm r, \bm x, t) \bm b(\bm r) \diff \bm r$, i.e., 
\begin{equation}
\begin{aligned}
&\partial_t \bm c_i + \bm u\cdot\nabla \bm c_i - \bm\kappa:\left\langle \sum_{j=1}^{N-1} \bm r_j\nabla_{\bm r_j}\otimes \bm b_i \right\rangle
\\
&= \frac{k_BT}{\gamma} \left\langle \sum_{j,k=1}^{N-1} A_{jk} \nabla_{\bm r_j} \cdot \nabla_{\bm r_k} \bm b_i \right\rangle  - \frac{1}{\gamma} \left\langle \sum_{j=1}^{N-1} \sum_{k=1}^{N_b} A_{jk} \nabla_{\bm r_k} V(\bm r) \cdot \nabla_{\bm r_j} \bm b_i \right\rangle,
\end{aligned}
\label{app_eq:FK_B_evoluation_multi}
\end{equation}
where $\bm A$ and $V$ specify the microscopic molecule bond connectivity structure and the interactions.  By comparing Eq. \eqref{app_eq:variational_PDE_form} and Eq. \eqref{app_eq:FK_B_evoluation_multi}, we can see that Eq. \eqref{app_eq:FK_B_evoluation_multi} provides a microscopic counterparts for the unclosed terms in Eq. \eqref{app_eq:variational_PDE_form}. 
Given a discrete sample of the polymer configuration $f(\bm r)$, we can evaluate these terms by the Monte-Carlo method without using the time-series samples. 

Crucially, the left-hand-side of Eq. \eqref{app_eq:FK_B_evoluation_multi} is frame-indifferent, and essentially, provides the microscopic counterpart for the macroscopic objective tensor derivative $\mathcal{D} \bm c_i /\mathcal{D} t$. In particular, it differs from the standard upper convective form due to the nonlinear form of encoder $\bm b_i$ and is reflected as the fourth-order tensor $\mathcal{E}_i$ in Eq. \eqref{app_eq:variational_PDE_form}. By establishing the micro-macro correspondence, this provides a generalized form of the objective tensor derivative with clear physical interpretation. 

\begin{proposition}
The left-hand-side of Eq. \eqref{app_eq:FK_B_evoluation_multi} is frame-indifferent.   
\end{proposition}

\begin{proof}
By Eq. \eqref{app_eq:encoder_b}, encoders $\bm b_i$ take the general form $\bm b_i = \sum_{j,k=1}^{N-1} \bm b_{jk}$, where $\bm b_{jk} = z(\bm r^{\ast}) \bm r_j {\bm r_k}^\top$ and $z$ is a scalar function. We consider the dynamics in two frames where frame 2 is rotating with respect to frame 1 with $\mathcal{U}(t)$, i.e., 
\begin{equation}
\frac{\rm d}{\rm d t} \left\langle \tilde{\bm b}_{jk}\right\rangle \big\vert_{\rm frame~1} = 
\dot{\mathcal U} \left\langle \bm b_{jk} \right\rangle \mathcal U^T
+ \mathcal U \left\langle \bm b_{jk} \right\rangle \dot{\mathcal U}^T
+ \mathcal U \frac{\rm d}{\rm dt} \left\langle \bm b_{jk} \right\rangle \big\vert_{\rm frame~2}
\mathcal U^T.
\label{app_eq:FK_B_1_angle}
\end{equation}
Furthermore, we have 
\begin{equation}
\begin{split}
&\tilde{\bm\kappa}:\left(\sum_{i=1}^{N-1}
\tilde{\bm r_i}\nabla_{\tilde{\bm r_i}}\otimes 
\tilde{\bm b}_{jk}\right) \big\vert_{\rm frame ~1} 
= \sum_{i=1}^{N-1} \left[ \left(\mathcal U \bm\kappa^\top \mathcal U^\top + \dot{\mathcal U} \mathcal{U}^\top\right) \cdot \mathcal U \bm r_i \right] 
\cdot \mathcal U \nabla_{\bm r_i} \left(\mathcal U \bm b_{jk}\mathcal U^\top\right) \\
&= \sum_{i=1}^{N-1} \mathcal U \left( (\bm\kappa^\top \cdot\bm r_i) \cdot \nabla_{\bm r_i} \bm b_{jk} \right) \mathcal U^\top
+ \mathcal U\left(\mathcal U^\top\dot{\mathcal U} \bm b_{jk} 
+ \bm b_{jk} \dot{\mathcal U}^\top \mathcal{U} \right)\mathcal U^\top \\
&= \sum_{i=1}^{N-1} \mathcal U \left( (\bm\kappa^\top \cdot\bm r_i) \cdot \nabla_{\bm r_i} \bm b_{jk} \right) \mathcal U^\top 
+ \dot{\mathcal U} \bm b_{jk} \mathcal U^\top
+ \mathcal U \bm b_{jk} \dot{\mathcal U}^\top.
\end{split}
\label{app_eq:FK_B_2_angle}
\end{equation}
By combining Eqs.~\eqref{app_eq:FK_B_1_angle} and \eqref{app_eq:FK_B_2_angle}, we conclude that 
\begin{equation}
\left. \left(\frac{\rm d}{\rm dt} \left\langle \tilde{\bm  b}\right\rangle 
- \tilde{\bm\kappa}:\sum_{i=1}^{N-1}\left\langle \tilde{\bm  r_i}\nabla_{\tilde{\bm  r_i}}
\otimes
\tilde{\bm  b}\right\rangle\right) \right\vert_{\rm frame~1} 
\equiv 
\mathcal{U} \left . \left[
\frac{\rm d}{\rm dt} \left\langle \bm  b \right\rangle 
- \bm\kappa:\sum_{i=1}^{N-1}\left\langle \bm  r_i \nabla_{\bm  r_i} \otimes
\bm  b \right\rangle \right] \right\vert_{\rm frame~ 2} \mathcal{U}^{\top}.
\label{app_eq:symmetry_Dc_Dt}
\end{equation}
\end{proof}

We remark that Eq. \eqref{app_eq:FK_B_evoluation_multi} provides the dynamics of $\bm c_i$ based on the over-damping Langevin dynamics with constant friction coefficient $\gamma$. We have computed the velocity autocorrelation functions of the polymer bead and verified that this free-draining model is accurate for the present microscopic system and the hydrodynamic interactions is smaller similar to Ref. \cite{jendrejack_stochastic_2002}. Eq. \eqref{app_eq:FK_B_evoluation_multi} can be further extended to incorporate the hydrodynamic interactions \cite{Zimm_model_JCP_1956, Bird_Curtiss_book_vol_2}, by which can be pass the microscopic interactions onto the macroscopic model. 

\subsection{Learning of VAR-Law (our model)}

With the micro-macro correspondence established between Eqs. \eqref{app_eq:variational_PDE_form} and \eqref{app_eq:FK_B_evoluation_multi}, we construct the present (VAR-Law) model through the joint learning of both the micro--macro encoders $\{\bm b_i(\bm r)\}_{i=1}^{N_c}$ and the variational ingredients $s$, $\mathcal E$,
and $\boldsymbol{\mathcal M}$. For fair comparison with the FENE-P model, we fix the microscopic definition of $\bm c_i$ as $\bm c_1
\propto \left\langle
\left(\sum_{k=1}^{N-1}\bm r_k \right) \left(\sum_{k=1}^{N-1}\bm r_k \right)^\top
\right\rangle$
throughout training where the scaling is conducted such that $\bm c_1^{eq} = \bm I$.  We learn the remaining conformation tensors $\bm c_2, \cdots, \bm c_{N_c}$ through the micro--macro encoders $\bm b_2, \cdots, \bm b_{N_c}$.

The macroscopic stress, relaxation, and flow-coupling terms are generated by the variational relations in
Eq.~\eqref{app_eq:variational_PDE_form} and matched to their
analytical microscopic counterparts. The loss is
\begin{equation}
\mathcal{L} = \sum_{l=1}^{N_s} \bigg( \Vert \bm H(\bm c^{(l)}) - \bm H^{{\rm micro},(l)}\Vert^2 +  \Vert \tilde{\mathcal{E}}(\bm c^{(l)}) - \tilde{\mathcal{E}}^{{\rm micro},(l)}\Vert^2 + \Vert \tau_p(\bm c^{(l)}) - \tau_p^{{\rm micro},(l)}\Vert^2 \bigg),   
\label{app_eq:law_learning_loss}
\end{equation}
where $l$ represents the discrete sample sets of the polymer configuration. The macroscopic terms and their microscopic counterparts take the form as 
\begin{equation*}
\begin{split}
&\tilde{\mathcal{E}}_i^{\rm macro} (\bm c_1, \cdots, \bm c_{N_c}) \qquad\qquad \tilde{\mathcal{E}}_i^{\rm micro} = 
\left\langle \sum_{j=1}^{N-1} \bm r_j \otimes \nabla_{\bm r_j}\otimes \bm b_i \right\rangle \\
&\bm H_i^{\rm macro} (\bm c_1, \cdots, \bm c_{N_c}) = \sum_{j=1}^{N_c} \mathcal{M}_{ij} :\frac{\partial s}{\partial \bm{c}_{j}} \\
&\bm H_i^{\rm micro} = \frac{k_BT}{\gamma} \left\langle \sum_{j,k=1}^{N-1}  A_{jk} \nabla_{\bm r_j} \cdot \nabla_{\bm r_k} \bm b_i \right\rangle - \frac{1}{\gamma} 
\left\langle \sum_{j=1}^{N-1} \sum_{k=1}^{N_b}  A_{jk} \nabla_{\bm r_k} V(\bm r) \cdot \nabla_{\bm r_j} 
\bm b_i \right\rangle, \\
&\bm\tau_p^{\rm macro}(\bm c_1, \cdots, \bm c_{N_c})= - \sum_{i=1}^{N_c} \tilde{\mathcal{E}}_{i}: \frac{\partial s}{\partial \bm c_{i}} \qquad \bm\tau_p^{\rm micro} =
n_p
\left\langle
\sum_{n=1}^{N_b}
\bm r_n\otimes\nabla_{\bm r_n}V(\bm r) 
\right\rangle - n_p N_b k_BT \bm I.
\end{split}    
\end{equation*}
Although encoder $\bm b_i(\bm r)$ do not explicitly appear in the macroscopic model terms in Eq. \eqref{app_eq:law_learning_loss}. They are still involved in the training through the construction of $\bm c_i$, as well as establishing the microscopic counterparts. 

Besides matching the deterministic stress and constitutive dynamics, the present joint learning framework can directly incorporate the friction operator $\bm{\mathcal{M}}$ similar to Ref. \cite{huang_statistical-physics-informed_2025}.  In particular, we can match $\bm{\mathcal{M}}$ with its microscopic counterparts by $\mathcal M_{pq} = (2k_BT/V_c)^{-1}
\langle\sum_{m_1,m_2,n}A_{m_1m_2}
\nabla_{m_1n}\bm b_p\otimes\nabla_{m_2n}\bm b_q\rangle$, where the covariance can be sampled from the fluctuation terms of the micro-model simulations. We tested the learning by incorporating this term, which yields models with similar accuracy.   

\subsection{Comparison models}

We compare VAR-Law with FENE-P and three additional learned
models: PDE-Law, VAR, and PDE. For the learned comparison
models, the encoders obtained from VAR-Law are reused and
held fixed. Thus, these models use the same conformation tensor variables $\bm c_1, \cdots, \bm c_{N_c}$, and only their macroscopic constitutive functions
are trained.

\paragraph{FENE-P model.}
The FENE-P model uses the prescribed tensor $\bm c_1$ (up to a scaling), denoted $\bm c$,
and the constitutive equations
\begin{equation}
\begin{split}
\bm\tau_p
&=
n_pk_BT(h\bm c-\bm I),\\
\partial_t\bm c+\bm u\cdot\nabla\bm c
&=
\bm\kappa^\top\bm c+\bm c\bm\kappa
+\frac{1}{\tau}(\bm I-h\bm c),\\
h
&=
\frac{b}{b+3-\operatorname{Tr}\bm c},
\end{split}
\label{app_eq:fene_p_model}
\end{equation}
where $b$ and $\tau$ are chosen for the best fit of the constitutive dynamics predicted from the microscopic model. The entropy density for FENE-P model is given as 
\begin{equation*}
s_\text{FENE-P}(\bm c)= \frac12 n_p k_B T
\left[ \log\det(\bm c)-b\log(h) \right]
\end{equation*}

\paragraph{PDE-Law model.}
The ``PDE-Law'' model retains the microscopic-law-based training targets but focuses on learning the macroscopic model directly in PDE form. Accordingly, the variational structure is not guaranteed. The conformation dynamics takes the form
\begin{equation}
\partial_t\bm c_i+\bm u\cdot\nabla\bm c_i
=
\bm\kappa:\tilde{\mathcal E}_i(\bm c)
+\bm H_i(\bm c).
\label{app_eq:pde_law_model}
\end{equation}
The stress $\bm\tau_p(\bm c)$, flow-coupling tensor
$\tilde{\mathcal E}_i(\bm c)$, and relaxation
$\bm H_i(\bm c)$ are parameterized independently similar to Ref. \cite{Lei_Wu_E_2020}. Rotational equivariance is enforced through an eigendecomposition-based representation. For fair comparison, these functions are trained using
Eq.~\eqref{app_eq:law_learning_loss}, with the fixed encoders $\bm c_i$ (the same as the ``VAR-LAW'' model)
used to evaluate the macroscopic states and microscopic
targets.

\paragraph{VAR model.}
The ``VAR'' model retains the macroscopic variational structure and trains the model by matching the evolution using macroscopic trajectory-based sample generated by microscopic simulations. Since the learning does not use explicit form of microscopic laws, we use the standard upper-convected transport.Its constitutive equations are
\begin{equation}
\begin{split}
\partial_t\bm c_i+\bm u\cdot\nabla\bm c_i
={}&
\bm\kappa^\top\bm c_i+\bm c_i\bm\kappa
+\bm H_i(\bm c),\\
\bm H_i(\bm c)
={}&
\sum_{j=1}^{N_c}
\mathcal M_{ij}(\bm c):
\frac{\partial s}{\partial\bm c_j},\\
\bm\tau_p(\bm c)
={}&
-2\sum_{j=1}^{N_c}
\bm c_j\frac{\partial s}{\partial\bm c_j}.
\end{split}
\label{app_eq:var_model}
\end{equation}
The entropy and mobility use the structure-preserving
representations described above.

The trajectory-based target is obtained by
subtracting the prescribed transport from the estimated
 evolution, i.e., 
\begin{equation}
\begin{split}
\bm H_i^{\rm traj}(t)
={}
\frac{\bm c_i(t+\Delta t)-\bm c_i(t)}{\Delta t}
+\bm u(t)\cdot\nabla\bm c_i(t)
-\bm\kappa(t)^\top\bm c_i(t)
-\bm c_i(t)\bm\kappa(t).
\end{split}
\label{app_eq:trajectory_target}
\end{equation}
The training loss is
\begin{equation}
\begin{split}
\mathcal L_{\rm traj}
=
\frac{1}{N_s}
\sum_{l=1}^{N_s}
\Bigg[
\left\|
\bm\tau_p(\bm c^{(l)})
-\bm\tau_p^{{\rm micro},(l)}
\right\|^2
+
\sum_{i=1}^{N_c}
\left\|
\bm H_i(\bm c^{(l)})
-\bm H_i^{{\rm traj},(l)}
\right\|^2
\Bigg].
\end{split}
\label{app_eq:trajectory_loss}
\end{equation}

\paragraph{PDE model.}
The ``PDE'' model uses the same prescribed upper-convected transport as the ``VAR'' model,
but represents the stress and relaxation independently, i.e., 
\begin{equation}
\partial_t\bm c_i+\bm u\cdot\nabla\bm c_i
=
\bm\kappa^\top\bm c_i+\bm c_i\bm\kappa
+\bm H_i(\bm c).
\label{app_eq:pde_model}
\end{equation}
Here, $\bm\tau_p(\bm c)$ and $\bm H_i(\bm c)$ are polynomial
functions of the conformation tensors, without variational
constraints. They are trained using the trajectory targets
and loss in Eqs. \eqref{app_eq:trajectory_target}
and \eqref{app_eq:trajectory_loss}.
All macroscopic models are coupled to the same
incompressible momentum equation.

\subsection{Training data}
One of the unique features of the present  model is that the training only requires time-discrete samples. This enables us to efficiently collect the training set using a set of embarrassingly parallel small-size microscopic simulations.

Specifically, we conduct independent microscopic simulations of a polymer solution with $N_m=1920$ polymer molecules in a small domain under either shear or elongation flow with different flow rates. The conformation tensors and microscopic constitutive targets in Eq. \eqref{app_eq:law_learning_loss} are evaluated by averaging over the configurations within
each ensemble. For the shear flow, we impose the different shear rate $\dot{\gamma}$.  For the elongation flow, we impose the prescribed velocity gradient $\bm\kappa = \operatorname{diag}
\left(
\dot\epsilon_1,\,
-\dot\epsilon_1+\dot\epsilon_2,\,
-\dot\epsilon_2
\right) $
with different $\dot\epsilon_1$ and $\dot\epsilon_2$. The Weissenberg number takes values between 0.05 and 5.0. 

For the 7-bead chain- and star-shaped polymer solution system, we choose $\dot\gamma$ over the range $0 \le \dot{\gamma} \le 0.02$,
$\dot{\epsilon}_1$ over the range $0 \le \dot\epsilon_1 \le 1.6\times 10^{-3}$, and $\dot{\epsilon}_2$  over the range of $-1\times 10^{-4} \le \dot\epsilon_2 \le 1\times 10^{-4}$. $50 - 200$ samples are collected from $200$ independent simulations, with a total of $30000$ samples.  

For the 15-bead chain-shaped polymer solution system, we choose $\dot\gamma$ over the range $0 \le \dot{\gamma} \le 0.01$, $\dot{\epsilon}_1$ over the range $0 \le \dot\epsilon_1 \le 6\times 10^{-4}$, and $\epsilon_2$ over the range of $-2\times 10^{-5} \le \dot\epsilon_2 \le 2\times 10^{-5}$. $50 - 200$ samples are collected from $200$ independent simulations, with a total of $30000$ samples. 

For comparison, the same sets of independent simulations are also used to collect the time-series samples with the same size of the training set for the learning of the ``VAR'' and ``PDE'' models.  

\subsection{Training procedure}

For both the seven-bead chain and star models, the trainable
encoders and macroscopic constitutive ingredients are first
optimized jointly for $48000$ steps using the ADAM optimization \cite{Kingma_Ba_Adam_2015}. The learning rate is initialized at $10^{-3}$ and decays
exponentially at each training step, decreasing by a factor
of $0.95$ over each epoch. Each iteration uses a mini-batch of $50$ ensemble samples per step.
The encoders are then frozen, and $\mathcal E$, $\boldsymbol{\mathcal M}$ and $s$ are further optimized for $4000$ steps at a fixed learning rate $10^{-4}$. The training is performed on one A100 GPU.

\section{Structure-preserving properties of the learned model}\label{app:properties}
\begin{proposition}\label{app_prop:positivity_preserving}
Assume that $\psi_e(\bm c)$, $\mathcal E_i(\bm c)$, and
$\mathcal M_{ij}(\bm c)$ are $C^1$ on
$(\mathbb S_{++}^3)^{N_c}$ and admit continuous extensions to
$(\mathbb S_{+}^3)^{N_c}$. Suppose that 
$\boldsymbol{\mathcal M}$ is uniformly positive definite, i.e.,
$
\bm A: \bm {\mathcal M}:\bm A
\ge
\epsilon \|\bm A\|_F^2,    
$
for $\bm A\in (\mathbb S^3)^{N_c}$ and some $\epsilon>0$. If $\bm c_i(0)\succ0$ and, along a
material trajectory, the conformation tensors and velocity gradient remain
bounded on a finite interval $[0,T)$, then there exists $\delta_T>0$ such that
\begin{equation}
\bm c_i(t)\succeq\delta_T\bm I,
\qquad
0\le t<T,\quad i=1,\ldots,N_c.
\end{equation}
\end{proposition}

\begin{proof}
Along a material trajectory, the conformation dynamics take the form
\begin{gather*}
\frac{\diff \bm c}{\diff t} = \bm{\mathcal{E}}_{\kappa} + \bm{\mathcal{M}} : \frac{\partial s}{\partial \bm{c}},
\end{gather*}
where $\bm{\mathcal{E}}_{\kappa,i} = \bm \kappa : \tilde{\mathcal{E}}_{i}$. 
By the assumed uniform boundedness of the conformation tensors and velocity gradient, and the continuity of the neural-network representations in their polynomial inputs, there exists $K_T>0$ such that $\|\bm{\mathcal E}_\kappa\|_F\le K_T$. 

Using the uniform positive definiteness of \(\bm {\mathcal M}\), we have 
\[
\begin{aligned}
\frac{ds}{dt}
&=\frac{\partial s}{\partial\bm c}:
\left(
\bm{\mathcal E}_{\kappa}
+\boldsymbol{\mathcal M}:
\frac{\partial s}{\partial\bm c}
\right)\\
&\ge
-K_T\left\|\frac{\partial s}{\partial\bm c}\right\|_F
+\epsilon\left\|\frac{\partial s}{\partial\bm c}\right\|_F^2\\
&\ge-\frac{K_T^2}{4\epsilon},
\end{aligned}
\]
which gives
\[
s(\bm c(t))\ge
s(\bm c(0))-\frac{K_T^2T}{4\epsilon}
=:s_T.
\]

On the other hand, recall that the entropy density takes the form
\[
s(\bm c)
=-\psi_e(\bm c)
-\left.\frac{\partial\psi_s}{\partial\bm c}
\right|_{\bm c^{\rm eq}}:\bm c
+\sum_{i=1}^{N_c}\lambda_i\log\det\bm c_i,
\]
with $\lambda_i>0$. Since the conformation tensors remain bounded and $\psi_e$ admits a continuous extension to the positive-semidefinite cone, the first two terms of $s(\bm c)$ are uniformly bounded above by a constant \(\psi_T\). Therefore, we have 
\[
\sum_{i=1}^{N_c}\lambda_i\log\det\bm c_i(t)
\ge s_T-\psi_T.
\]

By the assumed boundedness of $\bm c_i(t)$, there exists $c_T\ge1$ such that $\bm c_i(t)\preceq c_T\bm I$. Consequently, $\log\det\bm c_j(t)\le3\log c_T$, and for each $\bm c_i$, 
\begin{equation*}
\lambda_i\log\det\bm c_i(t)
\geq  s_T-\psi_T
-
3\log c_T\sum_{j\neq i}\lambda_j.    
\end{equation*}
Therefore, we have 
\[
\det\bm c_i(t)
\geq
\exp\left(
\frac{
s_T-\psi_T
-3\log c_T\sum_{j\neq i}\lambda_j
}{\lambda_i}
\right).
\]
Since the other two eigenvalues are bounded above by $c_T$, every eigenvalue of $\bm{c}_i(t)$ is bounded below by $\det\bm c_i(t)/c_T^2$. Let
\[
\delta_T
=
\frac{1}{c_T^2}
\min_{1\leq i\leq N_c}
\exp\left(
\frac{
s_T-\psi_T
-3\log c_T\sum_{j\neq i}\lambda_j
}{\lambda_i}
\right)>0,
\]
we obtain
\[
\bm c_i(t)\succeq\delta_T\bm I,
\qquad
0\leq t<T,\quad i=1,\ldots,N_c.
\]    
\end{proof}

\section{Structure-preserving numerical discretization}\label{app:discretization}

The extendable variational and thermodynamic admissible structure of the PDE form in Eq.~\eqref{app_eq:variational_PDE_form} provides the foundation for developing energy-stable and positivity-preserving numerical discretizations. 
Here, we briefly outline how this structure enables energy-stable spatial and temporal discretizations using the scalar auxiliary variable method \cite{Shen_Xu_Yang_SIAM_Review_2019} and the log-transform \cite{FATTAL200523}. For simplicity, we set the external body force $\bm g = 0$. 

Following the SAV method \cite{Shen_Xu_Yang_SIAM_Review_2019},  we consider the configurational free-energy contribution
$
\mathcal S_1=-\int_\Omega s(\bm c)\,d\bm x,
$
and introduce an auxiliary variable $z(t)=\sqrt{\mathcal S_1+C}$, 
where $C$ is chosen such that $\mathcal S_1 +C>0$.
Denoting $\bm G_i=-\partial s/\partial\bm c_i$
and $r=z/\sqrt{\mathcal S_1+C}$,
the PDE form in Eq.~\eqref{app_eq:variational_PDE_form} can be rewritten as
\begin{equation}
\begin{split}
\nabla\cdot\bm u &=0,\\
\rho_0(\partial_t\bm u+\bm u\cdot\nabla\bm u)
&=-\nabla p+\eta\Delta\bm u
+\nabla\cdot\left(
r\sum_{i=1}^{N_c}\tilde{\mathcal E}_i:\bm G_i
\right),\\
\partial_t\bm c_i+\bm u\cdot\nabla\bm c_i
-\bm\kappa:\tilde{\mathcal E}_i
&=-r\sum_{j=1}^{N_c}\mathcal M_{ij}:\bm G_j,\\
\frac{\diff z}{\diff t}
&=\frac{1}{2\sqrt{\mathcal S_1+C}}
\sum_{i=1}^{N_c}\int_\Omega
\bm G_i:\partial_t\bm c_i\,\diff\bm x.
\end{split}
\label{app_eq:extended_SAV}
\end{equation}
where the free energy dissipation follows
\begin{equation*}
\begin{gathered}
\mathcal H_{\mathrm{SAV}}
=\frac{\rho_0}{2}\int|\mathbf u|^2\, \diff \mathbf x
+z^2-C
\\
\frac{d\mathcal H_{\mathrm{SAV}}}{dt}
={}-\frac{\eta}{2}\int
|\bm D \bm  u|^2\, \diff \bm  x
-r^2\sum_{i,j=1}^{N_c}
\int \bm  G_i:\mathcal M_{ij}:\bm  G_j\, \diff \bm  x \le 0,
\end{gathered}
\end{equation*}
Following Ref. \cite{Shen_Xu_Yang_SIAM_Review_2019}, we can represent the  nonlinear terms in Eq. \eqref{app_eq:extended_SAV} as linearly implicit forms such as  
$r^{n+1}\mathcal{M}_{ij}^{n} :\frac{\partial s}{\partial \bm{c}_{j}^n}$ where $r^{n+1} = \frac{z^{n+1}}{\sqrt{\mathcal S_1^n+C}}$ and  $z$ can be updated by linearly implicit step as
\[
z^{n+1}-z^n
=\frac{\Delta t}{2\sqrt{\mathcal S_1^n+C}}
\sum_i\left\langle
\bm G_i^n,\,
\bm\kappa^{n+1}:\tilde{\mathcal E}_i^n
-r^{n+1}\sum_{j=1}^{N_c}
\mathcal M_{ij}^n:\bm G_j^n
\right\rangle.
\]
where the advection contribution vanishes under periodic and incompressible flow. With consistent discretizations of the stress and conformation dynamics, the time discretization preserves the modified energy dissipation as $\mathcal{H}_{\rm SAV}^{n+1} - \mathcal{H}_{\rm SAV}^n \le 0$. 

To preserve the positive definiteness of the conformation tensors, we introduce the logarithmic variables $\bm\Psi_i=\log\bm c_i$, following the log-conformation approach \cite{FATTAL200523}. As established in Proposition \ref{app_prop:positivity_preserving}, the continuous conformation tensors remain positive definite under the stated assumptions, so the logarithmic transformation is well-defined. Denoting the complete non-transport conformation rate by $\bm R_i$, i.e., 
\begin{equation}
\partial_t \bm {c}_{i}
+ \bm u \cdot \nabla \bm {c}_{i}  =   \bm R_i    \quad \bm R_i :=  \bm \kappa : \tilde{\mathcal{E}}_{i}  + \sum_{j=1}^{N_c} \frac{z}{\sqrt{\mathcal S_1+C}} \mathcal{M}_{ij} :\frac{\partial s}{\partial \bm{c}_{j}}
\end{equation}
the constitutive dynamics takes the form as 
\begin{equation}
\partial_t \bm {\Psi}_{i}
+ \bm u \cdot \nabla \bm {\Psi}_{i}  = D \log_{\bm c_i} [\bm R_i], 
\label{app_eq:log_transform_dynamics}
\end{equation}
where $D\log_{\bm c_i}[\bm R_i]$ represents the Fr\'echet derivative of the matrix logarithm at the positive-definite tensor \(\bm c_i\), applied to the symmetric tensor \(\bm R_i\).  The conformation tensors can then be reconstructed by $ \bm c_i^{n+1}=\exp(\bm\Psi_i^{n+1})\succ0$
provided \(\bm\Psi_i^{n+1}\) remains finite and symmetric.
Coupling the SAV and logarithmic representations requires a compatible discretization of the auxiliary-variable and conformation updates to retain the discrete energy law while preserving positive definiteness. 
Other positivity-preserving method \cite{BALCI_Thomeses_Symm_JNNFM_2011} may also be applied.

\end{appendices}

\bibliographystyle{unsrtnat}
\bibliography{main}
\end{document}